\documentclass[12pt]{iopart}

\newcommand{\beq}{\begin{equation}}
\newcommand{\eeq}{\end{equation}}
\newcommand{\beqn}{\begin{eqnarray}}
\newcommand{\eeqn}{\end{eqnarray}}
\newcommand{\ve}[1]{\mbox{\boldmath $#1$}}
\newcommand{\half} {{1\over 2}}
\newcommand{\threehalves} {{3\over 2}}
\newcommand{\fivehalves} {{5\over 2}}

\newcommand \sgam {\sqrt{\gamma}}

\newcommand{\eqref}[1]{(\ref{#1})}

\def\eadnew#1#2{\address{#2 E-mail: \mailto{#1}}}

\usepackage{graphicx}
\usepackage{epsf}
\usepackage{longtable}
\usepackage{hyperref}
\usepackage{graphics,epsfig,placeins,subfigure,wrapfig}
\usepackage[usenames]{color}
\usepackage{soul}

\begin{document}
\title{IllinoisGRMHD: An Open-Source, User-Friendly GRMHD Code for Dynamical Spacetimes\hspace{0.5cm} }

\author{
  Zachariah B.~Etienne$^{1,*}$,
  Vasileios Paschalidis$^{2,\dag}$,
  Roland Haas$^{3,\ddag}$,
  Philipp M\"osta$^{4,\S}$,
  Stuart L.~Shapiro$^{5,6,\P}$}
\address{$^1$ Department of Mathematics, West Virginia University, Morgantown, WV 26506, USA}
\address{$^2$ Department of Physics, Princeton University, Princeton, NJ 08544, USA}
\address{$^3$ Max-Planck-Institut f\"{u}r Gravitationsphysik, Albert-Einstein-Institut, D-14476 Golm, Germany}
\address{$^4$ TAPIR, Mailcode 350-17, California Institute of Technology, Pasadena, CA 91125, USA}
\address{$^5$ Department of Physics, University of Illinois at Urbana-Champaign, Urbana, IL 61801}
\address{$^6$ Department of Astronomy and NCSA, University of Illinois at Urbana-Champaign, Urbana, IL 61801}
\eadnew{zbetienne@mail.wvu.edu}{$^{*}$}
\eadnew{vp16@princeton.edu}{$^{\dag}$}
\eadnew{rhaas@aei.mpg.de}{$^{\ddag}$}
\eadnew{pmoesta@tapir.caltech.edu}{$^{\S}$}
\eadnew{slshapir@illinois.edu}{$^{\P}$}

\begin{abstract}

In the extreme violence of merger and mass accretion, compact objects
like black holes and neutron stars are thought to launch some of the
most luminous outbursts of electromagnetic and gravitational wave
energy in the Universe. Modeling these systems realistically is a
central problem in theoretical astrophysics, but has proven extremely
challenging, requiring the development of numerical relativity codes
that solve Einstein's equations for the spacetime, coupled to the
equations of general relativistic (ideal) magnetohydrodynamics (GRMHD)
for the magnetized fluids. Over the past decade, the Illinois Numerical Relativity (ILNR)
Group's dynamical spacetime GRMHD code has proven itself as a robust
and reliable tool for theoretical modeling of such GRMHD
phenomena. However, the code was written
``by experts and for experts'' of the code, with a steep learning
curve that would severely hinder community adoption if it were
open-sourced. Here we present IllinoisGRMHD, which is an open-source,
highly-extensible rewrite of the original closed-source GRMHD code of the ILNR
Group. Reducing the learning curve was the primary focus of this
rewrite, with the goal of facilitating community involvement in the
code's use and development, as well as the minimization of human
effort in generating new science. IllinoisGRMHD also saves computer time, generating
roundoff-precision identical output to the original code on
adaptive-mesh grids, but nearly twice as fast at scales of hundreds to
thousands of cores. 
\end{abstract}


\maketitle

\section{Introduction \& Motivation}

Incident gravitational wave (GW) observations have the potential to
address some of the most important unsolved problems in
astronomy and theoretical
astrophysics. These include testing GR in the strong field regime,
determining the engine behind short-hard gamma-ray-bursts,
constraining the equation of state above nuclear densities, revealing
how compact binaries form and evolve, as well as uncovering the
distributions of black hole spins and masses, just to name a
few. However, it is well known that unless some
coincident electromagnetic (EM) counterpart from the GW source is
observed, GW interferometers alone may be unable to pinpoint the
source position on the sky, hindering parameter estimation
\cite{Nissanke2011,LIGO_EM_FOLLOWUP_PAPER,ShahSluysNelemans2012,FanMessengerHeng2014}. 
Moreover, EM signals carry additional and complementary information
about the source, lending potentially critical insights about the GW
source and its environment.

Thus detections of EM counterparts to GWs could be critically
important in this age of ``multimessenger'' astronomy, and not solely
when GWs are detected first. For example, it may be possible that an
EM signal {\it itself} would imply a GW source,
leading to targeted searches across the GW spectrum, from the nHz
band in the case of dual AGNs, to the kHz band in the case of
stellar-mass binaries and supernovae. Beyond coincident GW detections,
EM transients linked to strong-field, dynamical spacetime phenomena
may {\it themselves} greatly advance our understanding of black hole
(BH) accretion phenomena and matter at extreme densities.

However, without detailed theoretical models of EM counterparts to GW
observations, our interpretation of observed EM counterparts may be
severely limited. Constructing such models remains a central problem
in theoretical astrophysics, for two key reasons. First, observable
signals are often sensitive to fluid flows and gravitational fields
spanning many orders of magnitude in lengthscale and
timescale. Second, the equations governing the dynamics are highly
complex and nonlinear, requiring the evolution of the full set of
Einstein's equations of general relativity, coupled to the equations
of general relativistic magnetohydrodynamics (GRMHD). 

Thus numerical relativity codes capable of modeling multi-scale GRMHD
flows promise to not only provide key insights into these important
phenomena, but represent the starting point for more sophisticated
modeling that includes advanced EM and neutrino radiation transport. 
More than a decade ago, the Illinois Numerical Relativity
Group led by S.~L.~Shapiro
was among the first to develop a dynamical spacetime GRMHD
code~\cite{Duezetal2005} for uniform-resolution grids. Since then,
this GRMHD code (henceforth, OrigGRMHD) has been significantly
extended and improved. Its current version models
multi-scale GRMHD flows via an adaptive-mesh-refinement (AMR)
vector-potential formulation. By evolving the vector potential forward
in time instead of the magnetic fields directly, this formulation
guarantees the no-monopole constraint $\nabla \cdot \ve{B}=0$ is
satisfied over the entire numerical grid, {\it even when multi-scale
  magnetized fluid flows cross AMR grid boundaries}. Notably, this
formulation reduces to the standard, staggered
Flux-constrained-transport (FluxCT)~\cite{balsara1999staggered} scheme
on uniform-resolution numerical
grids~\cite{Etienneetal2010,NewEMGauge,Farrisetal2012}. 

OrigGRMHD's reputation for generating models that address key unsolved
problems in theoretical astrophysics has been built upon years of
hard-fought development, as there exists no standard, proven
algorithms for dynamical spacetime, multi-scale GRMHD modeling. Over
the past decade, the code has been used to model a number of
astrophysical scenarios, gleaning key new insights into these
systems. For example, OrigGRMHD has produced state-of-the-art
magnetized binary neutron stars (NSs)
\cite{LiuMagnetizedBNS,Paschalidis2012} and binary BH--NS
\cite{EtienneFirstBHNS,EtienneSecondBHNS,EtienneThirdBHNS,EtienneFourthBHNS,Paschalidis2014}
models. It was also used to simulate magnetized disk accretion onto
binary BHs~\cite{Farrisetal2011,Farrisetal2012,Goldetal2014,Gold2014},
binary white dwarf--NS mergers~\cite{PaschalidisWDNS,Paschalidis2011},
magnetized, rotating NSs~\cite{Etienneetal2006}, magnetized Bondi
accretion~\cite{Etienneetal2010}, and magnetized hypermassive
NSs~\cite{HMNS6,HMNS1,HMNS2,HMNS3,HMNS4,HMNS5}, just to name a few.
The code was also recently extended, as a separate module, to solve
the equations of GR force-free electrodynamics \cite{Paschalidis2013b}
and applied to model both binary black hole--neutron star
\cite{Paschalidis2013a} and pulsar magnetospheres in GR
\cite{Ruiz2014}. At each stage of its development, OrigGRMHD was
subjected to a large battery of stringent test-bed
problems~\cite{Etienneetal2010}, which it had to pass before being
used for applications.

The field has matured considerably in the years since the first
dynamical spacetime GRMHD codes were announced, and multiple groups
now possess their own independent
codes~\cite{Duezetal2005,Anderson:2006ay,WhiskyMHD,CerdaDuran2008,Etienneetal2010,NewEMGauge,SACRA2012a,SACRA2012b,GRHYDRO,Dionysopoulou:2012zv,Palenzuela2013},
most of which solve these equations on adaptive mesh refinement (AMR)
grids. Given the
time and effort required to extend such codes to model more physics,
while still maintaining and improving the GRMHD modules, it seems
clear that the community might benefit if more of us consolidated our
efforts and adopted the same dynamic-spacetime GRMHD code.

With its proven robustness and reliability in modeling some of the
most extreme phenomena in the Universe, OrigGRMHD appears to be a
good candidate for such community adoption if it were
open-sourced. But despite its strong track record, OrigGRMHD was not
written with community adoption in mind, instead being a code written
``by experts and for experts'' of the code. As such, the code lacked a number of
features common to top-notch, widely-adopted open-source projects in
computational astrophysics, including sufficient documentation and
code comments, fine-grained modularity, a consistent coding style, and
regular, enforced code maintenance (e.g., removal of unused and
unmaintained features).

Thus the OrigGRMHD core development team came to the understanding
that unless these idiosyncrasies were fixed, open-sourcing the code
would be unlikely to engender widespread community adoption. Thus in
early 2013, it was decided to rewrite OrigGRMHD from the ground up,
with a focus on the four core design principles of user-friendliness,
robustness, modularity/extensibility, and
performance/scalability. Slightly more than a year later, all of
OrigGRMHD's core algorithms had been rewritten and the new code,
IllinoisGRMHD, was released.

Just after the decision was made to rewrite OrigGRMHD in 2013,
the first open-source, dynamical spacetime GRMHD code, called GRHydro,
was released~\cite{GRHYDRO}. Originally forked from the
dynamical spacetime, general relativistic {\it hydrodynamics} (GRHD)
Whisky code~\cite{Whisky} (not to be confused with its closed-source
successor, WhiskyMHD~\cite{WhiskyMHD}), GRHydro shares many of the
same features of OrigGRMHD, including a number of reconstruction
techniques. 

However, unlike OrigGRMHD/IllinoisGRMHD, GRHydro's GRMHD scheme has
not been developed to forbid the generation of monopoles (i.e.,
violations of the $\nabla \cdot \ve{B}=0$ constraint) when magnetized
fluids flow across AMR grid boundaries. As accurate modeling of such
multi-scale fluid flows is critically important in many astrophysical
scenarios of interest to the community, GRHydro's adoption by members
of the community has been limited, primarily to those who simulate core
collapse.

Further, one of OrigGRMHD/IllinoisGRMHD's key advantages is that these
codes are capable of stably modeling GRMHD flows into black hole (BH)
horizons over very long timescales, without the need for special
algorithms that excise GRMHD data within the BH. By contrast, it seems
that BH excision is an essential ingredient for stable GRMHD
evolutions with GRHydro in the presence of black holes. To date GRHydro
has been mostly used for core collapse (to a neutron star) simulations,
in which no black hole is present.

OrigGRMHD/IllinoisGRMHD have been demonstrated robust across a much
wider range of long-term BH simulations, and manage to do so without
excision. We conclude that making GRHydro's GRMHD schemes as robust
may require careful specification of boundary conditions on the
excision surface coupled to an interpolation scheme across AMR level
boundaries that respects the no-monopoles constraint,
e.g. \cite{balsara01,balsara09}.

IllinoisGRMHD was originally designed in a standalone sandbox to
maximize portability to other parallel infrastructures, but currently
adopts the latest Einstein Toolkit (ET)/Carpet AMR infrastructure. IllinoisGRMHD has been
proposed for inclusion within the next ET release, and code review is
underway. In the meantime, the ET community have graciously agreed to
host the IllinoisGRMHD code, in anticipation of official incorporation
upon completion of the code review process.\footnote{Instructions for
  downloading, compiling, and using IllinoisGRMHD may be found here: 
\url{http://math.wvu.edu/~zetienne/ILGRMHD/}}

In this paper, we present results from a number of code validation
tests demonstrating that IllinoisGRMHD (1) produces results identical
to OrigGRMHD, (2) possesses identical or significantly better
scalability and performance than OrigGRMHD and GRHydro,
(3) generates results in quantitative 
agreement with those of the GRHydro code, in the context of dynamical
spacetime evolutions of Tolman-Oppenheimer-Volkoff (TOV) stars.

The remainder of the paper is organized as
follows. Section~\ref{Basic_Equations} outlines the formulation of the
GRMHD equations solved by IllinoisGRMHD, Sec.~\ref{BasicAlgs} presents
a basic overview of algorithms used within IllinoisGRMHD,
Sec.~\ref{Validation} shows results from code validation tests,
Sec.~\ref{Bench} demonstrates the outstanding performance and
scalability of IllinoisGRMHD via benchmarks, and
Sec.~\ref{Conclusions} summarizes results and describes future work.

\section{Basic Equations}
\label{Basic_Equations}
All equations presented below are in geometrized units where $G=c=1$. In
these units, Einstein's equations become
\beq
\label{EinsteinsEquations}
G^{\mu\nu}=8\pi T^{\mu\nu}, \eeq 
where $G^{\mu\nu}$ is the Einstein tensor and $T^{\mu\nu}$ the
total stress-energy tensor. IllinoisGRMHD solves the coupled
Einstein-Maxwell equations assuming a perfect fluid stress-energy
tensor for the matter and infinite conductivity (ideal MHD), by
evolving via high-resolution-schock-capturing techniques the GRMHD
quantities that comprise the stress-energy tensor $T^{\mu\nu}$,
acting as the source for Einstein's equations.
With these assumptions, the GRMHD evolution and constraint equations
are derived from the following basic equations:
\begin{enumerate}
\item Conservation of baryon number
\beq
\nabla_\mu (\rho_0 u^\mu) = 0,
\label{restmass_conserv}
\eeq 
where $\nabla_\mu$ is the covariant derivative associated with the
spacetime metric tensor $g_{\mu\nu}$, $\rho_0$ is the fluid rest-mass
density and $u^\mu$ is the fluid four-velocity. 
\item Conservation of energy-momentum
\beq
\nabla_\mu T^{\mu\nu} = 0, T_{\mu\nu} = T^{\rm matter}_{\mu\nu} + T^{\rm EM}_{\mu\nu} 
\label{Emom_conserv}
\eeq
where $T_{\mu\nu}$ is the sum of the perfect fluid $T^{\rm
  matter}_{\mu\nu}$ and electromagnetic stress-energy tensors $T^{\rm
  EM}_{\mu\nu}$ in the ideal MHD limit ($u_\mu F^{\mu\nu}=0$).
\item Homogeneous Maxwell's equations
\beq
  \nabla_\nu F^{*\mu\nu} = \frac{1}{\sqrt{-g}} \partial_\nu (\sqrt{-g}
  F^{*\mu\nu}) = 0,
\label{Maxwells_Equations}
\eeq where $F^{\mu\nu}$ the Faraday tensor,
$F^{*\mu\nu}=(1/2)\epsilon^{\mu\nu\rho\sigma}F_{\rho\sigma}$ its dual
($\epsilon^{\mu\nu\rho\sigma}$ is the Levi-Civita tensor), and $g$ the
determinant of $g_{\mu\nu}$.
\end{enumerate}

As written, Eqs.~\eqref{restmass_conserv}, \eqref{Emom_conserv}, and
\eqref{Maxwells_Equations} for the plasma, as well as
Eq.~\eqref{EinsteinsEquations} for the spacetime metric, are not
particularly well-suited for numerical evolutions, so we choose
special formulations of them. For the spacetime metric evolution, we
choose an initial value formulation built upon first splitting the
4-metric $g_{\mu\nu}$ into the standard 3+1 Arnowitt-Deser-Misner
(ADM) form \cite{ADM}:
\beq ds^2 = g_{\mu\nu}dx^\mu dx^\nu= -\alpha^2 dt^2 +
\gamma_{ij} (dx^i + \beta^i dt) (dx^j + \beta^j dt).  
\eeq 
Here, $\alpha$ is the lapse function, $\beta^i$ the shift vector, and
$\gamma_{ij}$ the three-metric on spacelike hypersurfaces of constant
time $t$. This basic decomposition of the 4-metric can be used to
split the Einstein equations~\eqref{EinsteinsEquations} into a set of
evolution equations and a set of constraint equations that the
dynamical variables must satisfy for all times---similar to Maxwell's
equations---with projections of $T^{\mu\nu}$ along and normal to the 3D
spatial hypersurface existing as source terms (see, e.g.,
\cite{Baumgarte:2010nu} for a detailed discussion and
references). This original formulation of the Einstein equations is
known as the ADM 3+1 formulation of GR. A number of 3+1 formulations
can be derived from the ADM formulation and are useful for solving the
Cauchy problem for the Einstein equations. For the purposes of this
paper, we choose the Baumgarte-Shapiro-Shibata-Nakamura (BSSN)
formulation~\cite{SN,BS}, which introduces an auxiliary dynamical
variable and conformal scalings for the
dynamical variables, casting the evolution equations in a form that
allows for stable, long-term and accurate numerical integration.

To update $T^{\mu\nu}$ from one time slice to the next in a
simulation, IllinoisGRMHD solves Eqs.~\eqref{restmass_conserv},
\eqref{Emom_conserv}, and the spatial component of
Eq.~\eqref{Maxwells_Equations} in the ideal MHD limit ($u_\mu
F^{\mu\nu}=0$), as written in {\it conservative form}
(see e.g. \cite{Duezetal2005}):
\beq \partial_t \ve{C} + \ve{\nabla}\cdot \ve{F}
= \ve{S} \ ,
\label{eq:coneq}
\eeq
where $\ve{F}$ is the flux vector, $\ve{C}=\{\rho_{\star},
\tilde{\tau},\tilde{S_i},\tilde{B^i}\}$ the vector of conservative
variables, and $\ve{S}$ the vector of source terms. These vectors
depend directly on the ``primitive'' variables
$\ve{P}=\{\rho_0,P,v^i,B^i\}$, where $P$ is the pressure,
$v^i=u^i/u^0$ the fluid three-velocity, and $B^i$ are the spatial
components of the magnetic field ($B^\mu$) measured by normal (or
Eulerian) observers with four velocity $n^\mu=(1,-\beta^i)/\alpha$
(and is normal to the spatial hypersurface, $B^{\mu} n_{\mu}=0$, as well as
the metric and its derivatives. In particular, $\ve{C}$ may be written
in terms of $\ve{P}$, $\alpha$ and the metric as follows:
\beq
\ve{C} =
\left[
  \begin{array}{c}
    \rho_{\star} \\
    \tilde{\tau} \\
    \tilde{S_i} \\
    \tilde{B^i} 
  \end{array}
\right]
  = 
\left[
  \begin{array}{c}
    \alpha \sgam \rho_0 u^0 \\
    \alpha^2 \sgam T^{00} - \rho_{\star}\\
    (\rho_\star h + \alpha u^0 \sgam b^2) u_i - \alpha \sgam b^0 b_i \\
    \sgam B^i,
  \end{array}
\right]
\label{conservs}
\eeq
where $\gamma$ is the determinant of the 3-metric $\gamma_{ij}$,
$h=1+\epsilon+P/\rho_0$ is the specific enthalpy, with $\epsilon$ the
specific internal energy, and $b^\mu=B^\mu_{(u)}/\sqrt{4\pi}$ with
$B^\mu_{(u)}$ the magnetic field field measured by an observer
comoving with the fluid. Here the total stress-energy tensor
$T^{\mu\nu}$ can be written as follows (see
\cite{Baumgarte:2010nu,Duezetal2005,dlss05} for further details and
derivations)

\beq
\label{stressenergyperffluid}
T^{\mu \nu} = (\rho_0 h +b^2) u^{\mu} u^{\nu} + \left( P + \frac{b^2}{2}
\right) g^{\mu \nu} - b^{\mu} b^{\nu}.
\eeq

Our choice of fluid 3-velocity $v^i=u^i/u^0$ as a primitive variable
is consistent with a number of GRMHD codes, such as
\cite{HARM3D,East:2011aa,SACRA2012b}. However, our $v^i$ differs from that of
other GRMHD codes (e.g.,
\cite{GRHYDRO,Baiotti03a,Baiotti04,WhiskyMHD}, just to name a
few) that have adopted the Valencia formalism~\cite{Font1997,Anton2006},
which adopts the fluid 3-velocity $v_{(n)}^i$ as measured by normal
observers (also referred to as the Eulerian 3-velocity), defined as:
\beqn 
u^a &=& \alpha u^0 (n^a + v_{(n)}^a) \Rightarrow \\
v_{(n)}^i  &=& \frac{u^i}{\alpha u^0} + \frac{\beta^i}{\alpha}.
\eeqn
Note that $v_{(n)}^\mu$ is orthogonal to the normal vector to the 3D
spatial hypersurface $v_{(n)}^\mu n_\mu=0$. In terms of the fluid
3-velocity used by IllinoisGRMHD $v^i$, $v_{(n)}^i$ can be written as
\beq 
v_{(n)}^i = \frac{1}{\alpha} ( v^i + \beta^i ).  
\eeq
Note that this difference in the 3-velocity variable may account for
some of the differences in numerical results observed between
IllinoisGRMHD and (the Valencia-based) GRHydro in
Sec.~\ref{Validation}, as Valencia-based codes reconstruct $v_{(n)}^i$
instead of $v^i$.

Writing the GRMHD evolution equations in conservative form offers a
number of numerical advantages. First, without the source terms 
${\bf S}$, it guarantees conservation of total rest mass 
($\int_V \rho_* d^3x$), energy ($\int_V \tilde\tau d^3x$), and
momentum ($\int_V \tilde S_i d^3x$) to roundoff error. When the source
terms are accounted for, total ADM mass and momentum are conserved to
within truncation error. Second, it enables us to attach easily an
approximate Riemann solver, yielding a state-of-the-art
high-resolution shock-capturing (HRSC) numerical scheme, designed in
part to minimize oscillations near shocks. Such oscillations are
generated by approximating the flux derivative across shocks by smooth
functions. Third, the conservative form, coupled to the HRSC scheme
guarantees the shock jump conditions (Rankine-Hugoniot) are
satisfied. From an empirical perspective, finite-volume conservative
formulations, as implemented in IllinoisGRMHD, have been shown
superior at handling ultrarelativistic flows when compared to advanced
artificial viscosity (AV) schemes \cite{Mart1991,AnninosFragile2003}, despite
the fact that AV schemes are typically superior in terms of ease of
implementation and computational efficiency.

A key ingredient in a robust GRMHD code is the proper treatment of
the magnetic induction equation, which is derived from the spatial
components of Eq.~\eqref{Maxwells_Equations}. In conservative form,
these may be written: 
\beq \partial_t \tilde{B^i} + \partial_j \left(v^j
\tilde{B}^i - v^i \tilde{B}^j\right) = 0.
\label{B_induction}
\eeq
If the induction equation is directly evaluated in a numerical code
without special techniques, numerical truncation errors that violate
the divergence-free or ``no-monopoles'' constraint
\beq
\partial_i \tilde{B}^i = 0
\label{no_monopole}
\eeq
will be generated. Note that this equation is simply the time
component of Eq.~\eqref{Maxwells_Equations}. Maintaining satisfaction
of this constraint as the magnetic fields are evolved forward in time
[through direct evaluation of Eq.~\eqref{B_induction}] happens to
be a nontrivial endeavor, {\it particularly} on AMR grids. Our solution
\cite{Etienneetal2010,NewEMGauge} is to evolve the magnetic 4-vector
potential $\mathcal{A}_\mu$ instead of the magnetic fields directly,
so that
\beqn
\mathcal{A}_\mu &=& \Phi n_\mu + A_\mu, {\rm and} \\
\tilde{B}^i &=& \tilde\epsilon^{ijk} \partial_j A_k,
\label{BfromA}
\eeqn
where $A_\mu$ is purely spatial ($A_\mu n^\mu=0$) and $\Phi$ is
the EM scalar potential. Here, $\tilde\epsilon^{ijk}$ is the standard
permutation symbol, equal to 1 (-1) if $ijk$ are an even (odd)
permutation of 123, and 0 if one or more indices are
identical. Special finite difference operators for the vector
potential are defined in IllinoisGRMHD so that the divergence of a
curl is zero to roundoff error, which implies that the divergence of
Eq.~\eqref{BfromA} is zero and Eq.~\eqref{no_monopole} is satisfied
automatically, {\it even on AMR grids}.

In terms of $A_i$, the induction equation~\eqref{B_induction}
becomes 
\beq
\partial_t A_i = \tilde\epsilon_{ijk} v^j \tilde{B}^k 
- \partial_i (\alpha \Phi - \beta^j A_j).
\label{A_induction}
\eeq
What remains is to choose an EM gauge, and IllinoisGRMHD chooses the
``generalized Lorenz gauge condition'' by default that was introduced by
the Illinois Relativity group in~\cite{Farrisetal2012}. The covariant
version of the condition is $\nabla_\mu \mathcal{A}^\mu = \xi n_\mu
\mathcal{A}^\mu$, where $\xi$ is a parameter with dimensions $1/\rm Length$,
chosen carefully so that the Courant-Friedrichs-Lewy (CFL) factor
remains satisfied. Typically $\xi$ is set to $1.5/\Delta t_{\rm max}$,
where $\Delta t_{\rm max}$ is the timestep of the coarsest refinement
level. This gauge choice therefore yields the additional evolution
equation
\beq
\partial_t [\sgam \Phi] + \partial_j (\alpha \sgam A^j - \beta^j [\sgam\Phi]) = -\xi\alpha \sqrt{\gamma}\Phi.
\label{EMGaugeEvolEq}
\eeq
Note that IllinoisGRMHD evolves not $\Phi$ but $\sqrt{\gamma}\Phi$ as
the EM gauge variable. 

With the exception of this purely gauge evolution
equation and the vector-potential induction equation, all other GRMHD
evolution equations are written in 
conservative form and are solved via a HRSC scheme, as described in
Sec.~\ref{BasicAlgs}. For completeness, the remaining set of evolution
equations evolved by IllinoisGRMHD are written in conservative form
[Eq.~\eqref{eq:coneq}] as follows 
\beq \partial_t \left[
  \begin{array}{c}
    \rho_{\star} \\
    \tilde{\tau} \\
    \tilde{S_i}
  \end{array} 
  \right] +
 \partial_j 
 \left[
  \begin{array}{c}
    \rho_{\star} v^j \\
    \alpha^2 \sqrt{\gamma}\, T^{0j} - \rho_{\star} v^j \\
    \alpha \sqrt{\gamma}\, T^j{}_i
  \end{array} 
  \right]
 = 
 \left[
  \begin{array}{c}
    0 \\
    s \\
    \frac{1}{2}\alpha\sqrt{\gamma}\, T^{\alpha\beta} g_{\alpha\beta , i}
  \end{array} 
  \right],
\label{Full_Evol_Eqs}
\eeq
where
\beq
s=\alpha \sqrt{\gamma}\,\left[(T^{00} \beta^i \beta^j + 2 T^{0i}\beta^j + T^{ij})K_{ij} - (T^{00}\beta^i+T^{0i})\partial_i \alpha\right],
\eeq
and $K_{ij}=-\pounds_n \gamma_{ij}/2$ is the extrinsic curvature,
where $\pounds_n$ designates the Lie derivative along the hypersurface
normal vector $n$ (see e.g. \cite{Baumgarte:2010nu} for more details).

Finally, to close the system of equations, the EOS of the matter must
be specified. IllinoisGRMHD currently implements a hybrid EOS of the
form~\cite{jzm93}
\beq
\label{GammalawEOS}
  P(\rho_0,\epsilon) = P_{\rm cold}(\rho_0) + (\Gamma_{\rm th}-1)
\rho_0 [\epsilon-\epsilon_{\rm cold}(\rho_0)] \ ,
\eeq
where $P_{\rm cold}$ and $\epsilon_{\rm cold}$ denote the cold component 
of $P$ and $\epsilon$ respectively, and $\Gamma_{\rm th}$ is a constant 
parameter which determines the conversion efficiency of kinetic 
to thermal energy at shocks. The function $\epsilon_{\rm
  cold}(\rho_0)$ is related to $P_{\rm cold}(\rho_0)$ by the first law
of thermodynamics,
\beq
  \epsilon_{\rm cold}(\rho_0) = \int \frac{P_{\rm cold}(\rho_0)}{\rho_0^2} d\rho_0 \ .
\eeq
All functions within IllinoisGRMHD
support piecewise-defined $P_{\rm cold}(\rho_0)$ (the so-called
``piecewise polytrope'' EOS) with up to nine different polytropic
indices, except for the conservatives-to-primitives solver, which
currently supports only one. 

In all code tests presented in this paper, the $\Gamma$-law EOS
$P=(\Gamma-1)\rho_0 \epsilon$ is adopted. This corresponds to setting
$P_{\rm cold} = (\Gamma-1)\rho_0 \epsilon_{\rm cold}$ in
Eq.~\eqref{GammalawEOS}, which is equivalent to $P_{\rm
  cold}=\kappa\rho_0^\Gamma$ (with constant $\kappa$), and
$\Gamma_{\rm th}=\Gamma$. In the absence of shocks and in the initial
data used for our tests, $\epsilon=\epsilon_{\rm cold}$ and $P=P_{\rm
  cold}$.

\subsection{Outer Boundary Conditions}

We apply outer boundary conditions to primitive variables $\rho_0$,
$P$, and $v^i$, which enforce a zero-derivative, ``copy'' boundary
condition of these quantities at the outer boundary, {\it except} when
this results in a positive incoming velocity from the outer
boundary. Hence we refer to these as ``outflow'' boundary
conditions. As our outer boundary exists as a rectangular box, if for
example $v^x < 0$ at a given point on the $x=x_{\rm max}$ boundary
plane after applying the zero-derivative boundary condition, we set
$v^x = 0$ at that point. Similarly, at a given point on the $x=x_{\rm
  min}$ boundary plane, if $v^x > 0$ after applying the flat outer
boundary condition, $v^x$ is set to zero. The same strategy is applied
to the velocities for all other outer boundary faces. 

We also apply outer boundary conditions to $A_\mu$, linearly
extrapolating values to the outer boundary. To avoid problems caused
by reflections of $A_\mu$ waves from these imperfect outer boundary
conditions, the unit-bearing (i.e., not dimensionless) $\xi$
parameter in the $A_i$ evolution equation is set to some nonzero,
positive value, typically $1.5/\Delta t_{\rm max}$, where $\Delta
t_{\rm max}$ is the timestep of the coarsest refinement level.

\section{Basic Algorithms}
\label{BasicAlgs}




Since its initial stages, one of the primary objectives driving the
development of IllinoisGRMHD has been to remove nonrobust algorithms
and obsolete code from the original GRMHD code of the Illinois group,
resulting in a reliable state-of-the-art piece of software that is
more compact and easier for beginners to learn and extend. To this
end, all of the obsolete code and functionality proven to be nonrobust
in typical dynamical spacetime evolutions has been stripped from the
code, keeping within IllinoisGRMHD only the set of algorithms used in
all of the Illinois group's latest GRMHD publications. Further, the core algorithms
themselves have been rewritten into a uniform coding standard, with
large amounts of duplicated functionality replaced with a small,
optimized library of functions. This section reviews the basic
algorithms that comprise IllinoisGRMHD.


Here the algorithms that comprise the basic components of
IllinoisGRMHD are introduced, in the order in which they are
called. At the beginning of the first timestep, the variables
$\{P,\rho_0,v^i,B^i,A_\mu,\Phi\}$ must be defined at every
gridpoint. The following outlines the basic steps in which these
variables are updated at all gridpoints, in preparation for the next
timestep. All updates are performed by IllinoisGRMHD unless otherwise
specified. 
\begin{enumerate}
\item First, the flux and RHS terms in Eq.~\eqref{EMGaugeEvolEq} and
  Eqs.~\eqref{Full_Evol_Eqs}, for the set of evolution variables
  $\ve{E}=\{\rho_*,\tilde{S}_i,\tilde{\tau},A_i,[\sgam\Phi]\}$, are
  evaluated. Three separate algorithms are employed in this step:
\begin{enumerate}
\item A HRSC evolution scheme is used to compute the flux terms of the
  $\rho_*$, $\tilde{S}_i$, and $\tilde{\tau}$ evolution equations, as
  defined in Eqs.~\eqref{Full_Evol_Eqs}). This scheme, as well as the
  technique used to compute the source terms related to spacetime
  curvature in these equations, is described in
  Sec.~\ref{rhoStau_evol}.
\item Unlike the other primitive variables, $A_i$ and $B^i$ are
  defined on {\it staggered} gridpoints. Further, our $A_i$ evolution
  scheme is constructed to produce {\it identical} output to the
  standard, staggered constrained transport scheme of
  \cite{EvansHawleyCT1988}. As detailed in Sec.~\ref{A_evol}, this
  makes the HRSC scheme for updating $A_i$ a bit more involved than
  the HRSC scheme for evolving the unstaggered densitized density,
  momentum, and energy variables.
\item The evolution of the (staggered) EM gauge quantity $[\sgam
  \Phi]$ is not based on a HRSC scheme, as this quantity generally
  does not exhibit sharp features in our simulations. Its evolution
  algorithm is summarized in Sec.~\ref{Psi6Phi_evol}.
\end{enumerate}
\item Time derivative data from all evolved GRMHD variables are then
  passed to the MoL (Method of Lines) thorn, which iteratively
  integrates the evolved variables forward in time. MoL is capable of
  managing a number of iterative, explicit time integration
  techniques, of which we typically choose the four-iteration
  Runge-Kutta fourth-order (RK4) scheme, both in IllinoisGRMHD and the
  chosen spacetime evolution thorn.
\item Evaluating the time derivatives of all evolved GRMHD variables
  requires three ghostzones at the outer boundary of each AMR
  grid. The ghostzones at the outermost boundary are filled at each
  RK4 iteration, using the outer boundary update procedure outlined in
  the next steps. However, ghostzones at each internal AMR grid
  boundary are allowed to accumulate until the end of the fourth RK4
  iteration. Since RK4 consists of four iterations, this yields a total
  of $3\times4=12$ AMR grid boundary ghostzones that must be filled at
  the end of each full RK4 timestep. To fill these 12 ghostzones for
  all evolved variables at the end of the fourth RK4 iteration,
  prolongation and restriction operators are applied, which
  interpolate between different levels of refinement in both space and
  time. Third-order, line-averaged Lagrange prolongation/restriction
  is performed on $A_i$ and $\sgam \Phi$, and fifth-order Lagrange
  prolongation/restriction is performed on all other GRMHD evolved
  variables.
\item Next, linear-extrapolation outer boundary conditions are applied
  to $A_i$ and $[\sgam \Phi]$, as described in
  Sec.~\ref{OBCs}. Then $B^i$ is computed from $A_i$ at {\it all}
  gridpoints, as described in Sec.~\ref{A_evol}.
\item The conservative variables $\rho_*$, $\tilde{S}_i$, $\tilde{\tau}$, and
  $\tilde{B}^i$ have at this point been updated at all needed
  gridpoints, except at the outer boundary. However, the primitive
  variables $\{P,\rho_0,v^i,B^i\}$ do not yet exist at any
  gridpoints. As these variables are required at the outset of the
  next iteration, a conservative-to-primitives solver is called next,
  which at its heart employs a Newton-Raphson-based root-finder to
  invert Eqs.~\eqref{conservs}, computing primitive variables at each
  gridpoint based on the conservative variables at that
  point. Additionally, there are a number of consistency checks
  applied both before and after this solver is called. The
  procedure is outlined in Sec.~\ref{C2P}.
\item Next, zero-slope, outflow outer boundary conditions are applied
  to the set of primitive variables $\{P,\rho_0,v^i\}$, as described
  in Sec.~\ref{OBCs}. After this step, all variables
  $\{P,\rho_0,v^i,B^i,A_\mu\}$ needed to repeat this process have been
  defined at all gridpoints, so to proceed to the next timestep or RK4
  iteration, we simply loop to (i)(a). Values for the conservatives at
  the outer boundary are not strictly required for the evolution, but
  are set anyway, based on the primitive variables, in case a
  diagnostics utility might require that conservatives be set at the
  outer boundary.
\end{enumerate}

To conclude this introduction to IllinoisGRMHD's basic algorithms,
Sec.~\ref{ETLinkage} describes how IllinoisGRMHD connects to the rest of
the ET and its spacetime metric evolution modules.

\subsection{Evolution of $\rho_*$, $\tilde{S}_i$, and $\tilde{\tau}$}
\label{rhoStau_evol}
To evolve the GRMHD variables, IllinoisGRMHD first evaluates the
source terms of the time evolution equations for
$\{\rho_*,\tilde{S}_i,\tilde{\tau}\}$ in
Eqs.~(\ref{Full_Evol_Eqs}). Derivatives of the spacetime fields appear in the
source terms, which are evaluated via standard, second-order (default)
or fourth-order finite differences. Next, the fluxes are computed via
a second-order finite-volume HRSC
scheme. Since point values of gridfunctions and their volume averages
are the same to second order, our finite-volume scheme will converge
at the same order as a finite-difference scheme. We choose a
finite-volume scheme, as it will enable us to more rapidly move to a
higher-order method in future releases of IllinoisGRMHD, following the
strategy outlined in~\cite{Etienneetal2010}.

Computation of the flux term $\ve{\nabla}\cdot \ve{F} = \partial_m
F^m$ in a given direction $i \in \{x,y,z\}$ is performed with our
second-order finite-volume scheme in two steps, as detailed
below. First, the {\bf Reconstruction Step} computes values for the
primitive variables at cell interfaces (between gridpoints) along
direction $i$. Then the {\bf Riemann Solver} solves the Riemann
problem via an inexpensive, approximate algorithm, ensuring the
conservative variable fluxes between gridpoints are appropriately
constructed along direction $i$, even in the presence of
discontinuities or shocks. Upon completing the Riemann solver step for
a given flux direction $i$, the process is repeated in the other two
directions until $\partial_m F^m$ has been evaluated and summed in all
three spatial dimensions $\{x,y,z\}$.

{\bf The Reconstruction Step for $\rho_*$, $\tilde{S}_i$, and $\tilde{\tau}$:}
IllinoisGRMHD employs the Piecewise Parabolic Method (PPM) \cite{PPM},
incorporating the original flattening and steepening procedures
to reconstruct $\ve{P}$ at the right ($\ve{P_R}$) and left
($\ve{P_L}$) sides of each grid zone interface, along direction $i\in
\{x,y,z\}$. The version of PPM used within IllinoisGRMHD is designed
to maintain third-order accuracy, except at discontinuities or shocks
and at local minima and maxima. As in the GRHydro code \cite{GRHYDRO}
the flattening procedure within PPM was simplified to decrease the
number of required ghostzones within PPM from four to three.

After PPM reconstruction evaluates $\ve{P_{R,L}}$ along a given
direction $i$ and the metric values have been interpolated to each
grid zone interface at fourth-order (default) or second-order
accuracy, the fluxes $\ve{F_{R,L}}$ are then immediately evaluated via
\eqref{conservs} and Eqs.~\eqref{Full_Evol_Eqs}.

Next, for appropriate handling of fluxes across a given cell interface,
the Riemann problem must be solved.

{\bf The Riemann Solver for $\rho_*$, $\tilde{S}_i$, and $\tilde{\tau}$:}

The first step in solving the Riemann problem along direction
$i\in\{x,y,z\}$ is to compute the maximum ($+$) and minimum ($-$)
characteristic speeds $c^{R,L}_{\pm}$ at each cell interface,
approximating the general GRMHD dispersion relation (Eq.~27
of~\cite{HARM}) with the following, simpler expression: 
\beq
\omega_{\rm cm}^2 = \left[v_{\rm A}^2 + c_{\rm s}^2 (1 - v_{\rm A}^2 )
  \right] k_{\rm cm}^2,
\label{approx_disp_relation}
\eeq
Here, $\omega_{\rm cm}=-k_{\mu} u^{\mu}$ is the frequency and 
$k_{\rm cm}^2=K_{\mu}K^{\mu}$ the wavenumber of an MHD wave mode in the
frame comoving with the fluid, where $K_{\mu}$ is defined as the
projection of the wave vector $k^{\nu}$ onto the direction normal to
$u^{\nu}$: $K_{\mu}=(g_{\mu\nu} + u_{\mu}u_{\nu})k^{\nu}$. $c_{\rm s}$
is the sound speed, and $v_{\rm A}$ is the Alfv\'en speed, given by
\beq
v_{\rm A} = \sqrt{\frac{b^2}{\rho_0 h + b^2}}.
\eeq

With these definitions, the approximate dispersion relation
[Eq.~\eqref{approx_disp_relation}] may then be solved along direction
$i$, noting that the wave vector along this direction in the comoving
frame is given by $k_\mu=(-\omega,k_j \delta^j_i)$ and the wave
(phase) velocity by $c_{\pm} = \omega/(k_j \delta^j_i)$. The
dispersion relation can then be written as a quadratic equation for
$c_{\pm}$: $a_1 c_{\pm}^2 + a_2 c_{\pm} + a_3 = 0$, with $a_i$
given by
\beq
\begin{array}{ccc}
a_1 &=& (1-v_0^2) (u^0)^2 - v_0^2 g^{00}, \\
a_2 &=& 2 v_0^2 g^{i0} - 2 u^i u^0 (1-v_0^2), \\
a_3 &=&(1-v_0^2) (u^i)^2 - v_0^2 g^{ii},
\end{array}
\eeq
and $v_0^2 = v_{\rm A}^2 + c_{\rm s}^2 (1 - v_{\rm A}^2)$.

Though it makes $c_{\pm}$ simple to compute, this dispersion relation
overestimates the maximum characteristic speeds by a factor $\leq 2$,
which has the net effect of making the code more dissipative. Though
additional dissipation may smear important physical features in our
GRMHD flows, it also acts to help stabilize evolutions. Note that this
approximate dispersion relation is widely used in multiple codes
within the GRMHD community (e.g., WhiskyMHD \cite{WhiskyMHD},
GRHydro \cite{GRHYDRO}, HARM3D \cite{HARM3D}).

Once the maximum and minimum speeds $c_{\pm}$ have been computed at
left and right faces, the standard Harten-Lax-van Leer (HLL),
approximate Riemann solver \cite{Harten83} is then applied to compute
fluxes for the three conservative variables $\ve{U} =
\{\rho_*,\tilde{\tau},\tilde{S_i}\}$: 
\beq F^{\rm HLL} = \frac{c^-
  F_{\rm R} + c^+ F_{\rm L} - c^+ c^- (U_{\rm R} - U_{\rm L})}{c^+ +
  c^-},
\label{std_HLL_flux_formula}
\eeq
where $c^{\pm} = \pm \max(0,c^{\rm R}_{\pm},c^{\rm L}_{\pm})$, and
$U_{\rm R,L}$ are the conservative variables $\ve{U}$ computed from
the right and left reconstructed primitive values $\ve{P}_{R,L}$,
respectively.

Upon computing the HLL flux at cell interfaces, the final step in
evaluating the flux terms in the evolution equations of $\rho_*$,
$\tilde{S}_i$, and $\tilde{\tau}$ [Eqs.~\eqref{Full_Evol_Eqs}] is to
differentiate the computed HLL flux terms along the same direction
in which they were evaluated. After computing the HLL flux in the
x-direction we calculate the $x$-derivative of the flux as:
\beq
(\partial_x F^x)_{i,j,k} = \frac{F^x_{{\rm HLL}\; i+\half,j,k} -
  F^x_{{\rm HLL}\; i-\half,j,k}}{\Delta x}.
\eeq
The remaining $y$ and $z$-terms in the $\partial_m F^m$ sum are added
to the sum as reconstruction proceeds along the $y$ and
$z$-directions, respectively.

As the source terms of Eqs.~\eqref{Full_Evol_Eqs} have already been
computed, to complete the evaluation of $\partial_t \rho_*$,
$\partial_t \tilde{S}_i$, and $\partial_t \tilde{\tau}$, all
components of the $\partial_m F^m$ sum are then subtracted from
the source terms. These data are then passed to the MoL (Method of
Lines) thorn, which is capable of managing a number of explicit
time-stepping techniques. Although MoL supports a total-variation
diminishing third-order Runge-Kutta time integrator, we typically
choose the Runge-Kutta fourth-order (RK4) scheme for all evolution
variables, both in IllinoisGRMHD and the chosen spacetime evolution
thorn, as we find that RK4 minimizes the total error when evolving
both the spacetime and the fluid.

In parallel with evaluating the flux and source terms for $\rho_*$,
$\tilde{S}_i$, and $\tilde{\tau}$, IllinoisGRMHD employs a
vector-potential-based constrained transport scheme to evolve the
magnetic fields, which is detailed in the next section.

\subsection{Vector-Potential-Based Constrained Transport Scheme}
\label{A_evol}
Constrained-transport schemes maintain $\ve{\nabla}\cdot \ve{B}=0$
through careful finite differencing of the magnetic induction equation
flux terms [Eq.~\eqref{B_induction}]. Such schemes have proven highly
robust in the context of strongly-curved spacetimes; in particular
those inhabited by at least one black hole. These schemes are most
commonly and most directly applied in the context of
uniform-resolution grids. However, their use with AMR grids can be
complicated, as maintaining the divergenceless constraint at
refinement {\it boundaries} requires that special interpolations be
performed during prolongation/restriction.  Such
prolongation/restriction operators have been
devised~\cite{balsara01,balsara09}, but must be fine-tuned to the
particular AMR implementation.

IllinoisGRMHD applies an alternative constrained-transport scheme,
introduced by~\cite{HLL2D}. In this scheme, the magnetic induction
equation~\eqref{B_induction} is recast as an evolution equation for
the magnetic vector potential [Eq.~\eqref{A_induction}]. This scheme
has two important advantages. First, it produces identical output to
the standard, staggered constrained-transport scheme on uniform
resolution grids and thus shares its robustness. Second, evolving the
vector potential enables us to use {\it any} interpolation scheme at
AMR refinement boundaries without introducing nonzero divergence to
the magnetic fields, so long as we compute $B^i$ from the interpolated
$A_i$.

The remainder of this section details the staggered constrained-transport
scheme adopted within IllinoisGRMHD. First, we define the staggerings
of individual gridfunctions and the {\bf computation of $B^m$ from
  $A_m$}. Then the technique of {\bf reconstruction}
[Eq.~\eqref{B_induction}] on staggered cell faces is outlined, and
finally the {\bf Riemann solver} is described.

{\bf Computation of $B^m$ from $A_m$:}

In employing the standard, staggered constrained transport scheme,
magnetic fields are defined at gridpoints that are staggered with
respect to other conservative variables, as specified in
Table~\ref{tab:staggeredBA}. Notice that $A_m$ is staggered so that
$B^m$ may be computed immediately from Eq.~\eqref{BfromA} using the
following finite difference representation, accurate to second order:
\beq
B^x_{i+\half,j,k} = 
\sgam_{i+\half,j,k} (\partial_y A_z)^{i+\half,j,k} -
\sgam_{i+\half,j,k} (\partial_z A_y)^{i+\half,j,k},
\label{BxfromA}
\eeq
where 
\beqn
\sgam_{i+\half,j,k} &=& \exp\left( 6 \times \frac{1}{2} [\phi_{i,j,k} +
  \phi_{i+1,j,k}] \right), \\
(\partial_y A_z)^{i+\half,j,k} &=& \frac{A_z^{i+\half,j+\half,k} - A_z^{i+\half,j-\half,k}}{\Delta y}, \; {\rm and} \\
(\partial_z A_y)^{i+\half,j,k} &=& \frac{A_y^{i+\half,j,k+\half} - A_y^{i+\half,j,k-\half}}{\Delta z}.
\eeqn
Here, $\phi=(1/12)\log\gamma$ is the BSSN conformal exponent. Using
Eq.~\eqref{BxfromA} as a template, $B^y$ and $B^z$ can be written via
straightforward permutation of vector indices $\{x,y,z\}$, accounting
for the appropriate staggerings. Our finite differencing scheme is
specified so that the divergence of a curl is identically zero to
roundoff error. In this way, $B^m$ is guaranteed to be divergenceless
at all but the outermost ghost-zones on any given refinement level, so
long as $A_m$ is computed at all points. As with the other
conservative variables, reconstruction of the flux terms for $A_m$
requires three ghostzones (as discussed in the next section), so the
prerequisite step of computing $B^m$ from $A_m$ adds an additional
ghostzone, bringing the total number to four. However, we have found
that application of a copy boundary condition on $B^m$ to the
outermost gridpoint on each refinement level, coupled to the use of
only three ghostzones, results in qualitatively identical results to
runs that use four ghostzones. We find this to be the case even in the
most stringent tests, such as a magnetized BH accretion disk crossing
multiple refinement boundaries, as in
\cite{Farrisetal2012,Goldetal2014,Gold2014}. Thus by default, we have
used 3 ghostzones in all GRMHD simulations.

\begin{table}
\begin{center}
\caption{Storage location on grid of the magnetic field $B^i$ and
  vector potential $\mathcal{A}_{\mu}$. Note that $\ve{P}$ is the vector of
  primitive variables $\{\rho_0,P,v^i\}$}
\begin{tabular}{cc}
\hline
  Variable(s) & storage location \\
\hline
  Metric terms, $\ve{P}$, $\rho_*$, $\tilde{S}_i$, $\tilde{\tau}$ & $(i,j,k)$ \\
  $B^x$, $\tilde{B}^x$ & $(i+\half,j,k)$ \\
  $B^y$, $\tilde{B}^y$ & $(i,j+\half,k)$ \\ 
  $B^z$, $\tilde{B}^z$ & $(i,j,k+\half)$ \\
  $A_x$ & $(i,j+\half,k+\half)$ \\
  $A_y$ & $(i+\half,j,k+\half)$ \\
  $A_z$ & $(i+\half,j+\half,k)$ \\
  $\sgam \Phi$ & $(i+\half,j+\half,k+\half)$ \\
\hline
\end{tabular}
\label{tab:staggeredBA}
\end{center}
\end{table}

{\bf Flux Reconstruction of the Induction Equation:}

Accounting for staggerings, the evolution equation for $A_z$
(dropping the EM gauge terms to focus on the flux term of the
induction equation) is given by
\beq \partial_t A_z^{i+\half,j+\half,k} = -{\cal
  E}^z_{i+\half,j+\half,k}, \eeq where \beq {\cal E}^z = -v^x
\tilde{B}^y + v^y \tilde{B}^x 
\eeq 
is the flux term in the standard magnetic induction
equation~\eqref{B_induction}. Following \cite{HLL2D}, we
compute this flux term to staggered cell faces for $A_z$ and then
evaluate the HLL flux generalized for staggered grids. As ${\cal E}^z$
does not appear within a derivative of the $A_z$ induction equation
(as it does in the $\tilde{B^z}$ induction equation), the flux is not
directly finite-differenced prior to passing the right-hand side of
$\partial_t A_z$ to the time-stepping routines. Instead, the spatial
finite difference is computed {\it after} each RK4 iteration when
$B^m$ is computed from $A_m$. Critically, the order in which spatial
and temporal derivatives are evaluated is the {\it only} difference
between the standard, staggered constrained-transport scheme and our
vector-potential based staggered constrained-transport scheme. And
since spatial and temporal derivative operators commute within
IllinoisGRMHD's current framework, both schemes are identical on
uniform meshes.

Returning to the evaluation of ${\cal E}^z$, recall that primitives
such as $v^i$ are defined at grid points $(i,j,k)$, so computing the
value ${\cal E}^z$ at $(i+\half,j+\half,k)$ requires two successive
one-dimensional reconstructions of $v^x$ and $v^y$: first in the $x$
or $y$-direction and then in the $y$ or $x$-direction,
respectively. $\tilde{B}^x$ and $\tilde{B}^y$ already exist on
staggered gridpoints (see Table~\ref{tab:staggeredBA}), requiring only
a single reconstruction in the $y$ and $x$ direction,
respectively. Reconstruction is handled via the same PPM scheme as
described in Sec.~\ref{rhoStau_evol}.

{\bf Approximate Riemann Solver for $A_i$:}
The standard HLL formula~\eqref{std_HLL_flux_formula} for ${\cal
  E}^z$, generalized to the appropriate staggered gridfunctions, is
given by:
\beqn ({\cal E}^z)^{\rm HLL} &=& \frac{c^+_x c^+_y {\cal E}^z_{\rm LL}
  + c_x^+ c_y^- {\cal E}^z_{\rm LR} + c_x^- c_y^- {\cal E}^z_{\rm RL}
  + c_x^- c_y^- {\cal E}^z_{\rm RR}}{(c_x^+ + c_x^-)(c_y^+ + c_y^-)} \nonumber
\\ && + \frac{c_x^+ c_x^-}{c_x^+ + c_x^-} ( \tilde{B}^y_{\rm
  R}-\tilde{B}^y_{\rm L}) - \frac{c_y^+ c_y^-}{c_y^+ + c_y^-} (
\tilde{B}^x_{\rm R}-\tilde{B}^x_{\rm L}) \label{eq:Ezhll}\ \ \ 
\eeqn
In the above formula, ${\cal E}^z_{\rm LR}$ denotes the reconstructed
left state in the $x$-direction and right state in the
$y$-direction. Other symbols involving ${\cal E}^z$ are interpreted in
the similar fashion. $\tilde{B}^y_{\rm R}$ ($\tilde{B}^y_{\rm L}$)
denotes the reconstructed right (left) state of $\tilde{B}^y$ in the
$x$-direction, and $\tilde{B}^x_{\rm R}$ ($\tilde{B}^x_{\rm L}$)
denotes the reconstructed right (left) state in the $y$-direction. The
$c^{\pm}_x$ and $c^{\pm}_y$ should be computed by taking the maximum
characteristic speed among the four reconstructed states. However, we
set them equal to the maximum over the two neighboring interface
values for simplicity, as suggested in~\cite{HLL2D}, using the
technique described in Sec.~\ref{rhoStau_evol} to estimate the
speeds. The formula for $({\cal E}^x)^{\rm HLL}$ is obtained from
Eq.~(\ref{eq:Ezhll}) by permuting the indices $z \rightarrow x$, $x
\rightarrow y$ and $y\rightarrow z$, whereas the formula for $({\cal
  E}^y)^{\rm HLL}$ is obtained from Eq.~(\ref{eq:Ezhll}) by permuting
the indices $z\rightarrow y$, $x \rightarrow z$ and $y \rightarrow x$.

\subsection{Evolution of the densitized EM scalar potential $[\sgam \Phi]$}
\label{Psi6Phi_evol}

Incorporating the staggering of the EM gauge variable $[\sgam\Phi]$
(as specified in Table~\ref{tab:staggeredBA}), the evolution 
equation for $[\sgam \Phi]$ (Eq.~\ref{EMGaugeEvolEq}) may be written:
\beq
\partial_t [\sgam \Phi]_{i+\half,j+\half,k+\half} = 
- {\underbrace {\textstyle \partial_m (\alpha \psi^2 \tilde{\gamma}^{mn} A_n) }_{{\rm Term\; (1)}}} 
+ {\underbrace {\textstyle \partial_m \left( \beta^m [\sgam \Phi]\right)}_{{\rm Term\; (2)}}}
- {\underbrace {\textstyle \xi \alpha [\sgam \Phi]}_{{\rm Term\; (3)}}} ,
\eeq
where $\psi$ is the standard BSSN conformal factor and the relations
$\gamma^{mn} = \psi^{-4} \tilde\gamma^{mn}$ and $\sgam=\psi^6$ have
been applied. The left-hand side of the equation is evaluated at
$(i+\half,j+\half,k+\half)$, yet $A_m$, $[\sgam \Phi]$, and metric
quantities on the right-hand side of this equation all possess
different staggerings. Thus special care must be taken so that the
derivatives on the right-hand side (RHS) of this equation are
evaluated at gridpoints $(i+\half,j+\half,k+\half)$. To accomplish
this, quantities within the RHS derivatives are first interpolated to
consistent points prior to evaluation of the derivatives. Note that
this is strategy differs from the evolution of other GRMHD variables,
in that no reconstruction is applied. Methods for computing these
terms on the RHS are as follows:

{\bf Term (1):} For the $x$-derivative, all quantities within the
derivative operator ($[\alpha \psi^2] \tilde{\gamma}^{mx} A_m$) are first
interpolated to $(i,j+\half,k+\half)$. At second-order accuracy,
interpolations to staggered gridpoints are trivial, requiring only
averages of neighboring unstaggered points. For example, interpolation
of $\alpha \psi^2$ from $(i,j,k)$ to $(i,j+\half,k+\half)$ is
performed in two steps:
\begin{enumerate}
\item $[\alpha \psi^2]_{i,j+\half,k} = \half ( [\alpha \psi^2]_{i,j,k} +[\alpha \psi^2]_{i,j+1,k})$
\item $[\alpha \psi^2]_{i,j+\half,k+\half} = \half ( [\alpha \psi^2]_{i,j+\half,k} +[\alpha \psi^2]_{i,j+\half,k+1})$
\end{enumerate}
Once $[\alpha \psi^2]$, $\tilde{\gamma}^{mx}$, $A_y$, and $A_z$ have
been interpolated in this way to $(i,j+\half,k+\half)$, the derivative
$\partial_x ([\alpha \psi^2]\tilde{\gamma}^{xm} A_m)$ is computed to
second order as follows 
\beq 
\partial_x ([\alpha \psi^2] \tilde{\gamma}^{xm} A_m)_{i+\half,j+\half,k+\half} = \frac{ ([\alpha
    \psi^2]\tilde{\gamma}^{xm}A_m)_{i+1,j+\half,k+\half} - ([\alpha
    \psi^2]\tilde{\gamma}^{xm}A_m)_{i,j+\half,k+\half} }{\Delta x}. 
\eeq 
Other derivatives in the sum $\partial_m(\alpha \psi^2
\tilde{\gamma}^{mn} A_n)$ are computed in the same fashion.

{\bf Term (2):} The computation of this term is made easier by the
fact that $[\sgam \Phi]$ is staggered at $(i+\half,j+\half,k+\half)$
already. So to evaluate the derivative, $\beta^m$ is first
interpolated from $(i,j,k)$ to $(i+\half,j+\half,k+\half)$ using the
same interpolation strategy as with {\bf Term (1)}. Next, notice that
this term is basically a shift advection term on the EM gauge quantity
$[\sgam \Phi]$. Such advection terms are typically upwinded within
the metric evolution thorn, so for consistency we apply the same
upwinding strategy when evaluating this derivative:
\beq
\partial_m \left(\beta^m [\sgam \Phi]\right) = \left\{ \begin{array}{ll}
  D^-_m \left(\beta^m [\sgam \Phi]\right) & \mbox{if $\beta^m < 0$\ ,} \\
  D^+_m \left(\beta^m [\sgam \Phi]\right) & \mbox{otherwise}
  \end{array}
\right.
\eeq
where the second-order operators are
\beq
(D^-_x f)_{i+\half,j+\half,k+\half} =
\frac{
  f_{i-\threehalves,j+\half,k+\half}
-4f_{i-\half,j+\half,k+\half}
+3f_{i+\half,j+\half,k+\half}}{2 \Delta x}
\eeq
and
\beq
(D^+_x f)_{i+\half,j+\half,k+\half} =
\frac{
 -f_{i+\fivehalves,j+\half,k+\half}
+4f_{i+\threehalves,j+\half,k+\half}
-3f_{i+\half,j+\half,k+\half}}{2 \Delta x}
\eeq
for the derivative in the $x$-direction. Derivatives in the $y$ and
$z$-directions follow in a straightforward fashion.

{\bf Term (3):} 
The computation of this term is also made easier by the fact that
$[\sgam \Phi]$ is staggered at $(i+\half,j+\half,k+\half)$ already. So
to evaluate it, only $\alpha$ must be interpolated from $(i,j,k)$ to
$(i+\half,j+\half,k+\half)$ using the same interpolation strategy as
with {\bf Term (1)}.

\subsection{Conservatives-to-Primitives Solver}
\label{C2P}
After the conservative GRMHD variables have been updated at all
gridpoints, with boundary conditions and prolongation/restriction
operators applied, the primitive variables must then be computed from
the conservative variables. This is not a trivial endeavor, as the
conservative variables generally depend on the primitive variables in
a nonlinear way, requiring the implementation of a root-finding
method. To this end, IllinoisGRMHD employs the two-dimensional
Newton-Raphson solver of \cite{Noble2006,HARM3D}.

Truncation errors originating from spatial and temporal finite
differencing, as well as interpolation, prolongation, and restriction
operations can push the evolved GRMHD quantities to unphysical values,
resulting in either unphysical values for the primitive variables or
no values at all. For definitions of unphysical values of the
GRMHD quantities please see the Appendix A
of~\cite{EtienneThirdBHNS}. So prior to calling
the two-dimensional Newton-Raphson solver, we perform a number of checks
that determine whether the conservative variables are in a physically
valid range. If they are not, they are modified prior to calling the
root-finder. Even with these checks, the Newton-Raphson solver will occasionally
fail to find a root. This is very rare, and almost always occurs in a
low-density atmosphere or inside a black hole. In such an instance, we
set the pressure to $P_{\rm cold}$, which guarantees a successful
inversion. The implementation of these checks and modifications have
been described in detail in Appendix A of~\cite{EtienneThirdBHNS}.

After the Newton-Raphson solver has successfully found a set of
primitives, the primitives are checked for physicality, and if they
are not in the physical range, they are minimally modified until they
return to the physical range. First, if the velocity is found to be
superluminal, the speed is reduced to IllinoisGRMHD's default Lorentz
factor limit, which is set to $W=10$, where $W$ is the Lorentz factor
of the fluid as measured by a normal observer. Next, IllinoisGRMHD
does not include any cooling mechanism, which means that for
evolutions adopting a $\Gamma$-law equations of state, the pressure
should not physically drop below $P_{\rm cold}$. So a pressure floor
of $0.9 P_{\rm cold}$ is imposed. Increasing this floor to $P_{\rm
  cold}$ exactly results in large central density drifts in TOV star
evolutions. Simulations can crash in the other extreme, if $P/P_{\rm
  cold}$ becomes too large. This typically only happens in very low
density regions or inside black holes. So at densities $\rho_0<100
\rho_{\rm atm}$ or deep inside black hole horizons, a ceiling on $P$
of $100 P_{\rm cold}$ is enforced (see Appendix A of
\cite{EtienneThirdBHNS} for more details).

\subsection{Outer Boundary Conditions for $A_i$, $[\sgam \Phi]$, $P$, $\rho_0$, and $v^i$}
\label{OBCs}
Updating evolved variables within IllinoisGRMHD requires three
ghostzones per RK4 iteration, and at the end of each iteration, outer
boundary conditions are applied to $A_i$, $[\sgam \Phi]$, $P$,
$\rho_0$, and $v^i$ fill these ghostzones. The algorithm applies the
outer boundary conditions in all directions, from the innermost
gridpoint outward, as follows. For example, in the positive
$x-$direction, the first outer gridpoint $i+1$ is defined as
\beq
\ve{E}_{i+1} = \left\{ \begin{array}{ll}
  \ve{E}_{i}, & \mbox{if $\ve{E} \in \{P,\rho_0,v^y,v^z\}$, or $\ve{E}\equiv v^x$ and $v^x\geq 0$} \\
  0, & \mbox{if $\ve{E}\equiv v^x$\ , and $v^x<0$} \\
  2 \ve{E}_{i} - \ve{E}_{i-1}, & \mbox{if $\ve{E}\in\{[\sgam \Phi],A_x,A_y,A_z\}$}
  \end{array}
\right.
\eeq
And for the negative $x-$direction, the first outer gridpoint $i-1$ is
defined as
\beq
\ve{E}_{i-1} = \left\{ \begin{array}{ll}
  \ve{E}_{i}, & \mbox{if $\ve{E} \in \{P,\rho_0,v^y,v^z\}$, or $\ve{E}\equiv v^x$ and $v^x\leq 0$} \\
  0, & \mbox{if $\ve{E}\equiv v^x$\ , and $v^x>0$} \\
  2 \ve{E}_{i} - \ve{E}_{i+1}, & \mbox{if $\ve{E}\in\{[\sgam \Phi],A_x,A_y,A_z\}$}
  \end{array}
\right.
\eeq
As for the positive/negative $y$ and $z$ directions, the procedure
is the same, replacing $v^x \leftrightarrow v^y$ and
$v^x\leftrightarrow v^z$, respectively.

In this way, linear extrapolation outer boundary conditions are
applied to the vector potential variables $\{[\sgam \Phi],A_i\}$ and
zero-derivative, outflow outer boundary
conditions are applied to the hydrodynamic variables
$\{P,\rho_0,v^i\}$. These conditions are applied to the innermost
gridpoints on the three-gridpoint-thick outer boundary surface, first
in the positive $x$, then $y$, then $z$ directions, followed by the
negative $x$, then $y$, then $z$ directions. Next, they are applied to
the second innermost gridpoints on the outer boundary surface in all
directions, and finally to the outermost point.

Currently, IllinoisGRMHD supports the use of the $xy-$plane as a
symmetry plane, in which case the negative $z-$direction outer
boundary condition is not applied, letting the Cactus/Carpet
parallel AMR infrastructure impose the reflection symmetry.

\subsection{Linkage of IllinoisGRMHD to the Rest of the Einstein Toolkit}
\label{ETLinkage}
In order to evolve the GRMHD equations in a dynamical spacetime
context, IllinoisGRMHD must be coupled to a separate module that
evolves the spacetime metric, typically using components of the
stress-energy tensor produced by IllinoisGRMHD as source terms. The ET
is based within the Cactus infrastructure, thus modules are called
``thorns'', of which IllinoisGRMHD is one. ET is structured so that
thorns evaluating the spacetime metric evolution equations (i.e., the
left-hand side of Einstein's equations) must couple to a common
interface thorn, called ADMBase. Similarly, thorns that evaluate
evolution equations governing the right-hand side of Einstein's
equations couple to an interface thorn called TmunuBase. TmunuBase and
ADMBase are designed to interface seamlessly, so that GRMHD evolution
thorns coupled to TmunuBase will automatically work with {\it any}
spacetime evolution thorn properly coupled to ADMBase. Thus, since
IllinoisGRMHD is fully coupled to TmunuBase, it is immediately
compatible with all spacetime evolution formulations within ET,
including BSSN, and conformal and covariant Z4
\cite{Alic:2011gg} (both provided by the McLachlan
\cite{McLachlan,Kranc:web} Thorn).  For dynamical spacetime evolutions
within this paper, IllinoisGRMHD is coupled to the McLachlan BSSN
thorn.

\section{Code Validation Tests}
\label{Validation}
This section compares the results of IllinoisGRMHD to those of two
other codes written using the ET infrastructure: GRHydro, which is the
only other open-source GRMHD code within ET, and the original,
closed-source GRMHD code of the Illinois group (OrigGRMHD), on which
IllinoisGRMHD is based. OrigGRMHD has been subjected to a large
battery of stringent test-bed problems, including but not limited to
standard 1D relativistic MHD shock tests, 2D cylindrical blast
explosion tests, magnetized Bondi accretion and stellar collapse
tests as well as (self-)convergence tests~\cite{Etienneetal2010}.
Though it may be argued that IllinoisGRMHD has not been as robustly
tested as GRHydro or OrigGRMHD, we demonstrate here that IllinoisGRMHD
and OrigGRMHD results agree to roundoff error, indicating that both
are algorithmically identical, and GRHydro and IllinoisGRMHD results
agree within truncation error, indicating that both can be expected to
converge to the same result.

The first two parts of this section (Secs.~\ref{TOVroundoff} and
\ref{randomIDroundoff}) demonstrate that IllinoisGRMHD and OrigGRMHD
generate output identical to roundoff error, in two complementary,
highly-challenging tests.  In the first test (Sec.~\ref{TOVroundoff}),
a weakly-magnetized TOV star is evolved over many dynamical timescales
and grid light-crossing times. The second test
(Sec.~\ref{randomIDroundoff}) exposes both codes to a type of ``fuzzing'',
in which random initial data are evolved. Unlike the first test,
initial data in the second test contain strong shocks,
highly-magnetized and highly-relativistic, stochastic flows, as well
as nontrivial, discontinuous spacetime-metric and extrinsic curvature
components. Despite the harshness of the second test, both codes are
shown in Sec.~\ref{randomIDroundoff} to produce roundoff-error
identical results over many grid light-crossing times. Results from
these tests are highly significant, as they demonstrate that
IllinoisGRMHD yields identical results to the ``battle-hardened'',
trusted code on which it is based.

In our final validation test, both IllinoisGRMHD and GRHydro evolve 
unmagnetized, stable TOV stars in a dynamical spacetime backdrop, and
are shown to converge to the same result at the expected order, though
IllinoisGRMHD exhibits slightly slower central density drift and lower
Hamiltonian constraint violations at a given resolution.

\subsection{IllinoisGRMHD and OrigGRMHD: Roundoff-Error Agreement in
  Evolving Magnetized TOV Star}
\label{TOVroundoff}

\begin{figure}
\raggedleft
\includegraphics[angle=270,width=0.8\textwidth]{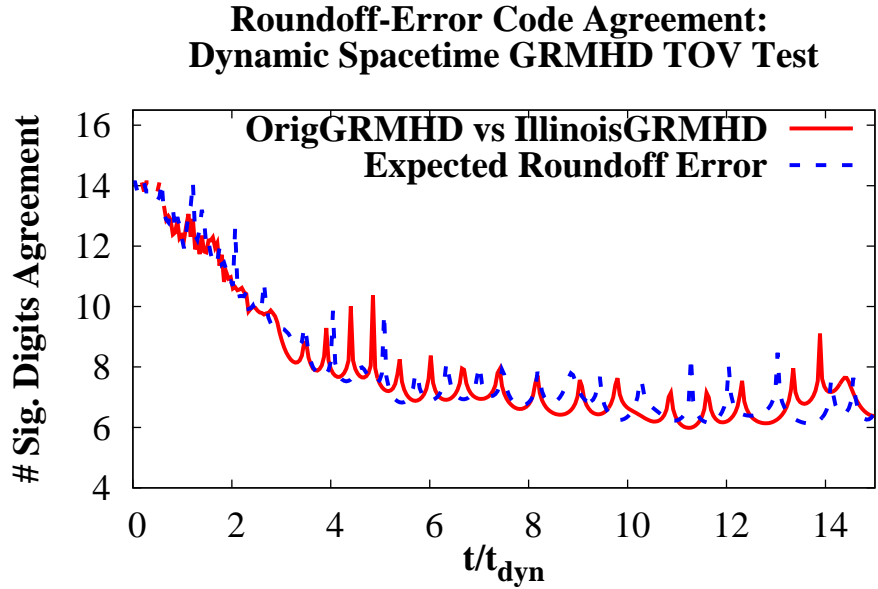}
\caption{Significant digits of agreement between pairs of codes,
  monitoring the central density of a magnetized neutron star in a
  dynamic-spacetime GRMHD simulation versus time, as measured in
  dynamical timescales $t_{\rm dyn}=1/\sqrt{\rho_{0,\rm max}}$. The solid red line
  shows the number of significant digits of agreement between
  IllinoisGRMHD and OrigGRMHD. The dashed blue line shows the expected
  roundoff error, measured as the number of significant digits of
  agreement between OrigGRMHD and itself with a 15th-significant-digit
  perturbation to the initial density of the magnetized neutron
  star. This run was performed on 8 parallel processes on a desktop
  computer. All details from this simulation are provided in
  \ref{App:Mag_roundoff_test_setup}.
}\label{roundoff_TOV_output}
\end{figure}

IllinoisGRMHD, OrigGRMHD, and GRHydro all implement double-precision,
64-bit floating point arithmetic, which represents numbers to
between 15--17 significant digits. Given the sophisticated and
iterative nature of these GRMHD codes, initial machine-precision
differences can grow enormously over time. As an example, we multiply
the initial rest-mass density of a weakly-magnetized TOV star by
$1+10^{-15}$, yielding a 15th significant digit perturbation. We then
perform the evolution, measuring the number of significant 
digits of agreement between this perturbed run and an unperturbed
evolution, through 15 dynamical timescales and on AMR grids with
multiple levels of refinement. The dashed blue line of
Fig.~\ref{roundoff_TOV_output} plots the result from this test. Notice
that the number of significant digits of agreement quickly drops from
14 digits, plateauing to between 6 and 8 digits of agreement.

We were careful to develop IllinoisGRMHD so that its results agree
with OrigGRMHD to roundoff error, and Fig.~\ref{roundoff_TOV_output}
confirms that for this weakly-magnetized TOV star test, the number of
significant digits of agreement between IllinoisGRMHD and OrigGRMHD
(solid red line) follow the same curve as the expected roundoff error
intrinsic to OrigGRMHD (dashed blue line). For full details of the
physical scenario modeled here, as well as the grid parameters, see
\ref{App:Mag_roundoff_test_setup}.

\subsection{IllinoisGRMHD and OrigGRMHD: Roundoff-Error Agreement in
  Evolving Random Initial Data}
\label{randomIDroundoff}

Although we have demonstrated that when evolving weakly-magnetized TOV
stars on AMR grids, IllinoisGRMHD and OrigGRMHD produce results that
agree to roundoff error, one might argue that even though there is
a sharp discontinuity at the stellar surface, this code test is
insufficient for truly demonstrating roundoff-level agreement, as it
lacks strong shocks and highly-relativistic, highly-magnetized fluid
flows. To address this potential criticism, we developed a random
initial data module that sets up both weak and strong stochastic GRMHD
flows atop an artificially-static, weak and strong-field stochastic
spacetime background that is nearly conformally flat. The stochastic
nature of these data means that both metric and GRMHD quantities
suffer from both weak and strong discontinuities from one
spatial point to the next, providing a robust test of the
high-resolution shock-capturing algorithms within these GRMHD codes,
as well as a confirmation of the stability of both GRMHD codes to fuzz
testing. We stress that although the chosen metric has Lorentzian
signature and the spatial three-metric is positive-definite, these
initial data are for numerical convenience only and are not designed
to satisfy Einstein's equations. In fact, when these
GRMHD data are evolved forward in time, spacetime field variables are
strictly held fixed in time.

All components of spacetime and GRMHD tensors and vectors are nonzero,
randomly fluctuating from one spatial point to the next, with each
component having a unique magnitude. As a result, this module has been
useful in checking for typos in the GRMHD evolution equations, which
were completely rewritten in IllinoisGRMHD. For example, if by mistake
$\gamma_{xy}$ were written $\gamma_{xz}$ in any of the GRMHD
equations, then because these components have differing magnitudes,
IllinoisGRMHD and the original GRMHD code of the Illinois group would
not agree to roundoff precision and the test would fail. This module
was used extensively in the first stages of IllinoisGRMHD
development to find such typos, as well as truncation-error-level
algorithmic differences between the old and new codes. All such typos
and algorithmic differences were fixed and modified, respectively, so
that roundoff-level agreement could be demonstrated. A full
description of the random initial data module and grid setup is
provided in \ref{App:random_id}.

\begin{figure}
\raggedleft
\includegraphics[angle=270,width=0.8\textwidth]{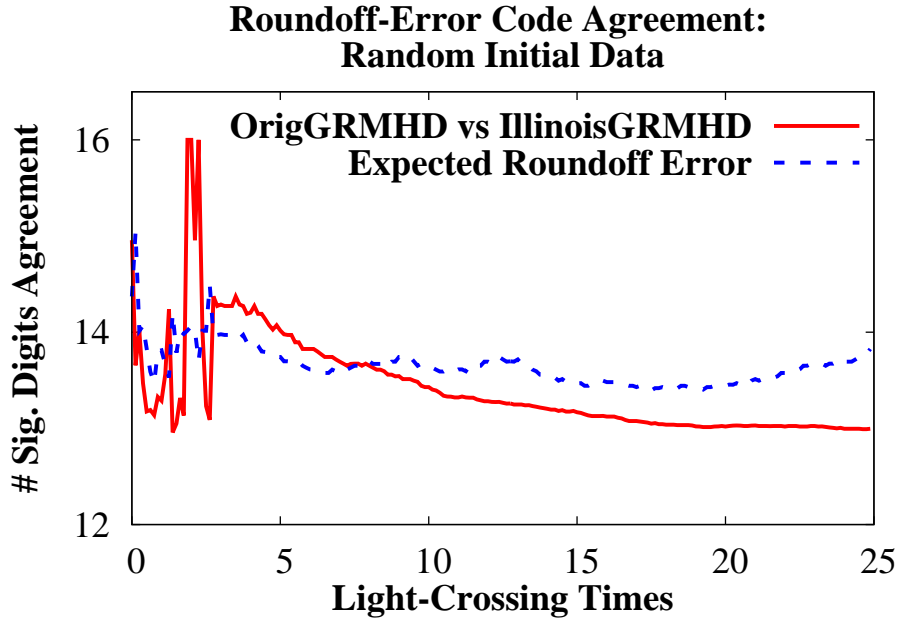}
\caption{Significant digits of agreement between pairs of codes,
monitoring the rest-mass density $\rho_0$ summed along the $x$-axis on
the finest AMR level. The solid red line shows the number of
significant digits of agreement between IllinoisGRMHD and
OrigGRMHD. The dashed blue line shows the expected roundoff error,
measured as the number of significant digits of  agreement between
OrigGRMHD and itself with a random 15th-significant-digit perturbation
to all primitive GRMHD variables. This run was performed on 8 parallel
processes on a desktop computer, and was shown to agree with the
single-process OrigGRMHD run to roundoff-error as well. All details
from this simulation are provided in
\ref{App:Mag_roundoff_test_setup}.}
\label{roundoff_random_id}
\end{figure}

As with the magnetized TOV roundoff-error test of
Sec.~\ref{TOVroundoff}, we measure the expected level of roundoff
error by first adding a random, 15th-digit perturbation to all GRMHD
primitive variables after they are set. Then we evolve both perturbed
and unperturbed initial data with the trusted OrigGRMHD code. The
difference grows with time, but plateaus to about 13 digits over time,
as shown in Fig.~\ref{roundoff_random_id}. 

Next, the same unperturbed initial data are evolved on the same grids
with IllinoisGRMHD, and the results confirm that IllinoisGRMHD and
OrigGRMHD indeed agree to within expected roundoff error, through at
least 25 light-crossing times. All of these runs were performed on 8
parallel processes. Full details regarding how these initial data are
generated, as well as the computational setup for these simulations,
are provided in \ref{App:Mag_roundoff_test_setup}.

\subsection{IllinoisGRMHD and GRHydro: Unmagnetized TOV Star
  Convergence Tests}
\label{TOVConvergence}

The open-source IllinoisGRMHD and closed-source OrigGRMHD have been
shown to produce roundoff-error identical results even when evolving
very harsh, relativistic, strongly-magnetized, discontinuous initial
data. IllinoisGRMHD represents the second open-source,
dynamical spacetime GRMHD module, the first being GRHydro \cite{GRHYDRO}. Both
are based in the Einstein Toolkit, which provides a particularly
convenient infrastructure for performing GRMHD simulations in a
dynamical spacetime context. GRHydro contains a large number of
features, including a variety of reconstruction options, approximate
Riemann solvers,  and outer boundary options. OrigGRMHD contains many
such features as well, but nearly all of these features are not robust
in the context of black-hole-inhabited spacetimes and have thus
remained unused for years. IllinoisGRMHD contains only the features
from OrigGRMHD that have been used in all recent papers by the
Illinois NR group (e.g., \cite{EtienneThirdBHNS,EtienneFourthBHNS,Farrisetal2012,Goldetal2014,Gold2014,Paschalidis2014}).

This section compares results between IllinoisGRMHD and the
standard, FORTRAN version of GRHydro, using identical initial data,
computational grids, dynamical spacetime evolution (BSSN) modules,
reconstruction scheme, and Riemann solver. Since IllinoisGRMHD has
been shown to agree with OrigGRMHD to roundoff precision, these tests
can also be seen as a proxy comparison between GRHydro and OrigGRMHD.

As detailed in~\ref{App:Unmag_conv_test_setup}, both IllinoisGRMHD and
GRHydro evolve the same physical scenario in this test as in
Sec.~\ref{TOVroundoff}, but with the magnetic fields inside the TOV
star set to zero. GRHydro options were chosen so that its algorithms
would be identical to IllinoisGRMHD. Despite the basic algorithms
being the same, both codes differ significantly in how they
are implemented. This difference in implementation {\it should} result
in truncation-level differences between the two codes, but instead we
find slightly different convergence properties between the codes.

A quantity $Q(\Delta x)$ that converges to zero at $n$th
order with increasing resolution (i.e., decreasing grid spacing
$\Delta x$) satisfies
\beq
\frac{Q(\Delta x_1)}{Q(\Delta x_2)} = \left(\frac{\Delta x_1}{\Delta x_2}\right)^n.
\eeq
Thus the convergence order to zero, $n$, is written as follows:
\beq
n = \log\left( \frac{Q(\Delta x_1)}{Q(\Delta x_2)} \right) \left / \log
\left( \frac{\Delta x_1}{\Delta x_2} \right) \right..
\label{ndef}
\eeq

\begin{figure}
\includegraphics[angle=270,width=0.5\textwidth]{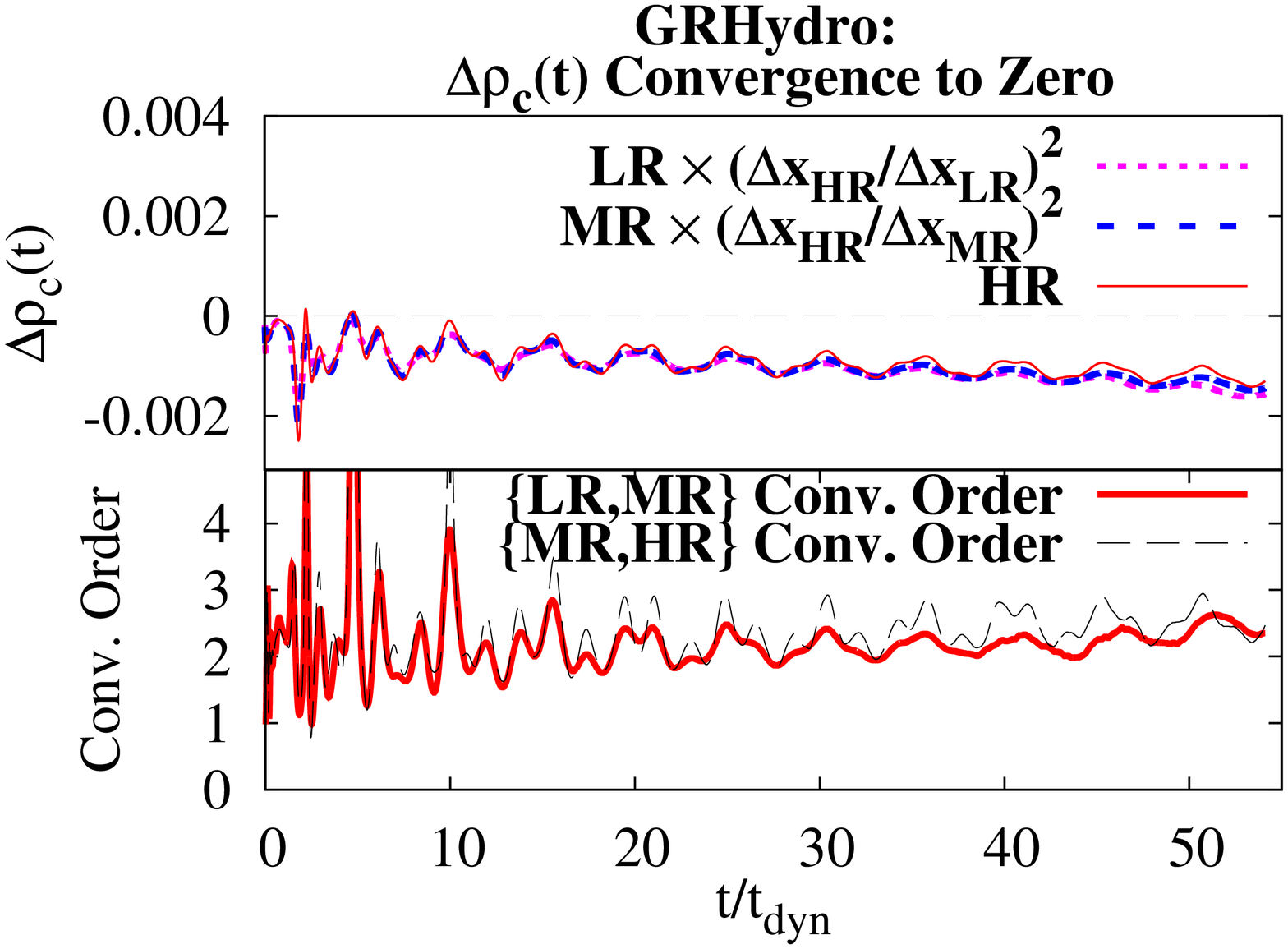}
\includegraphics[angle=270,width=0.5\textwidth]{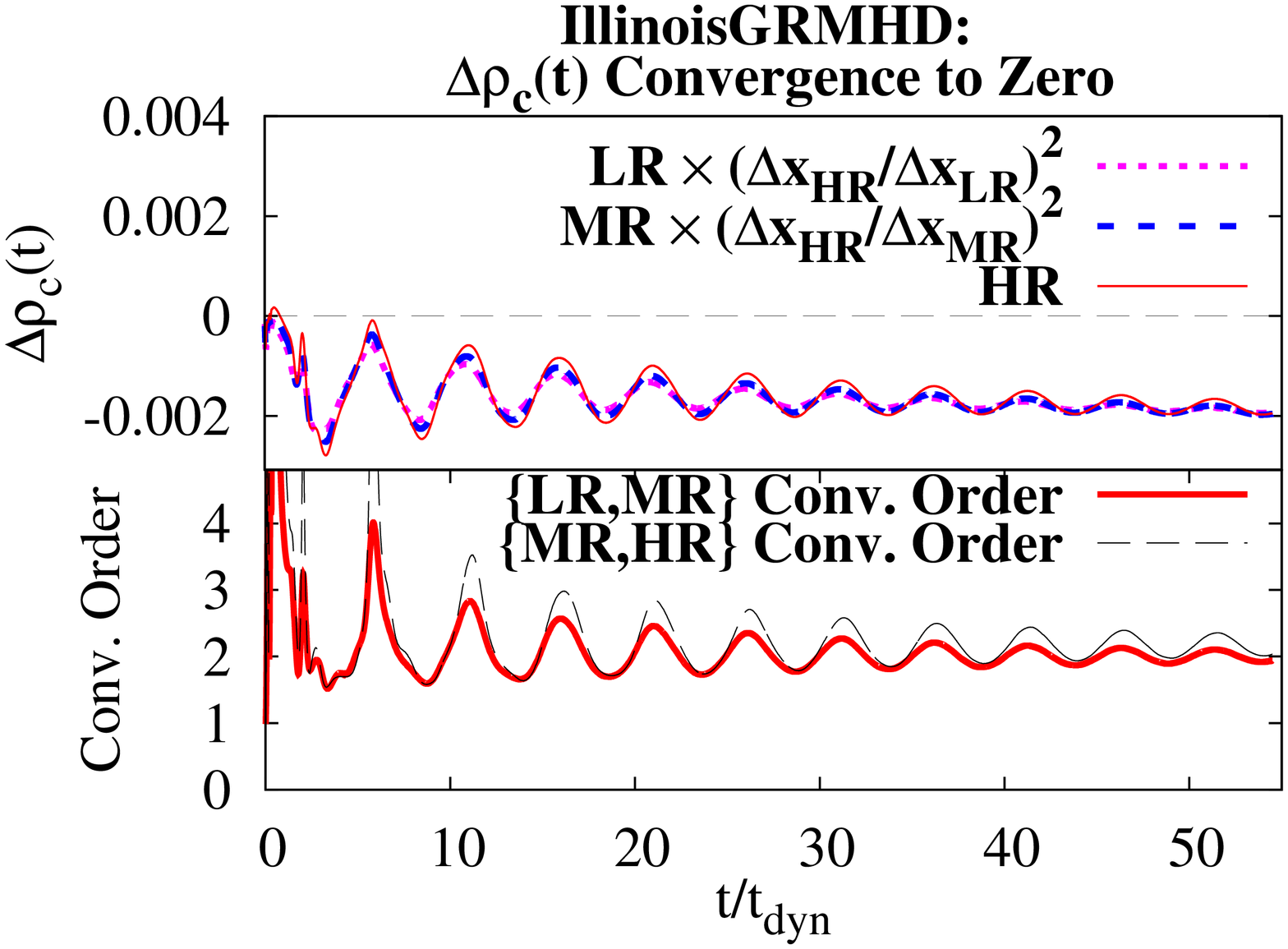}
\\
\includegraphics[angle=270,width=0.5\textwidth]{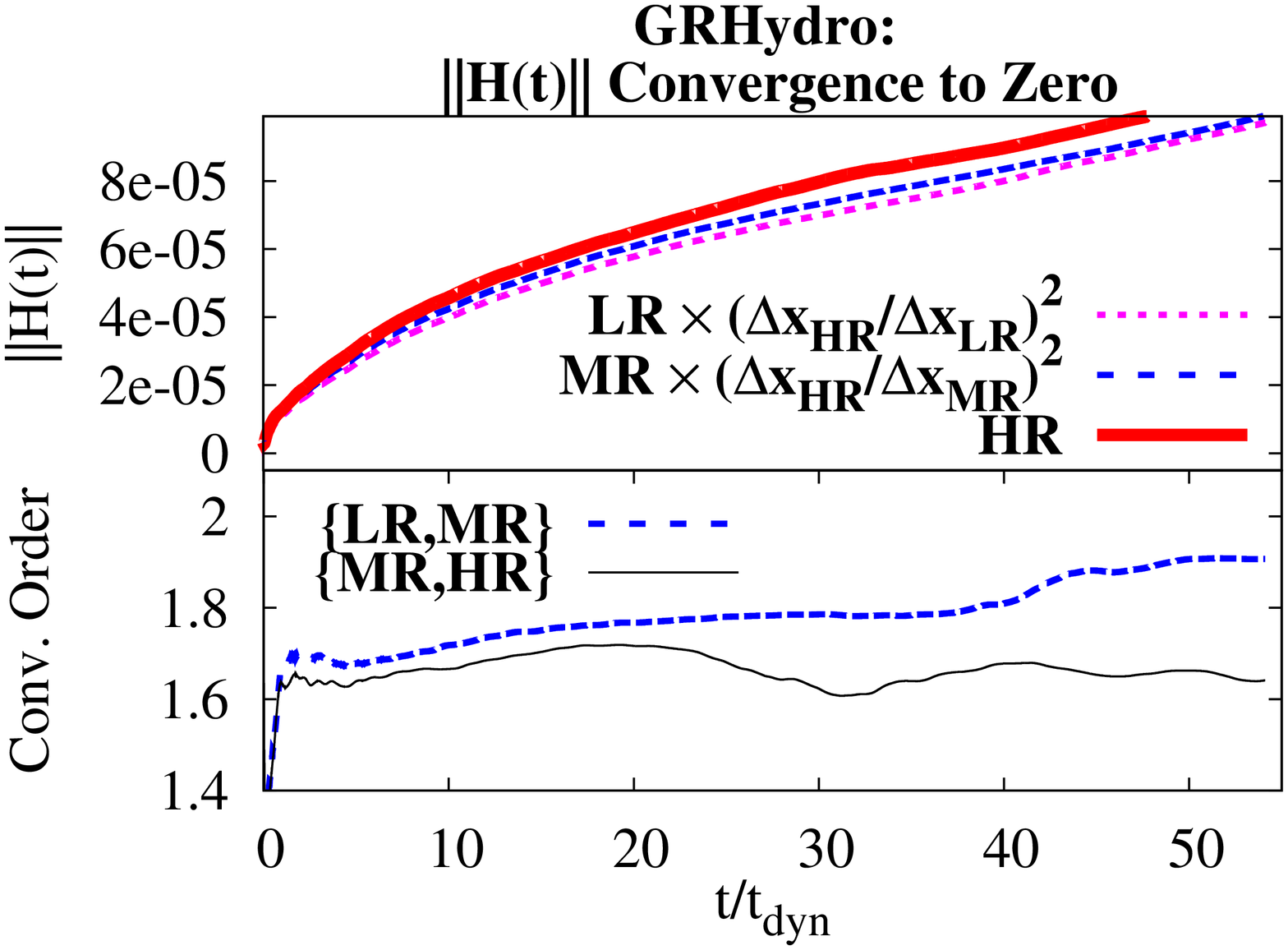}
\includegraphics[angle=270,width=0.5\textwidth]{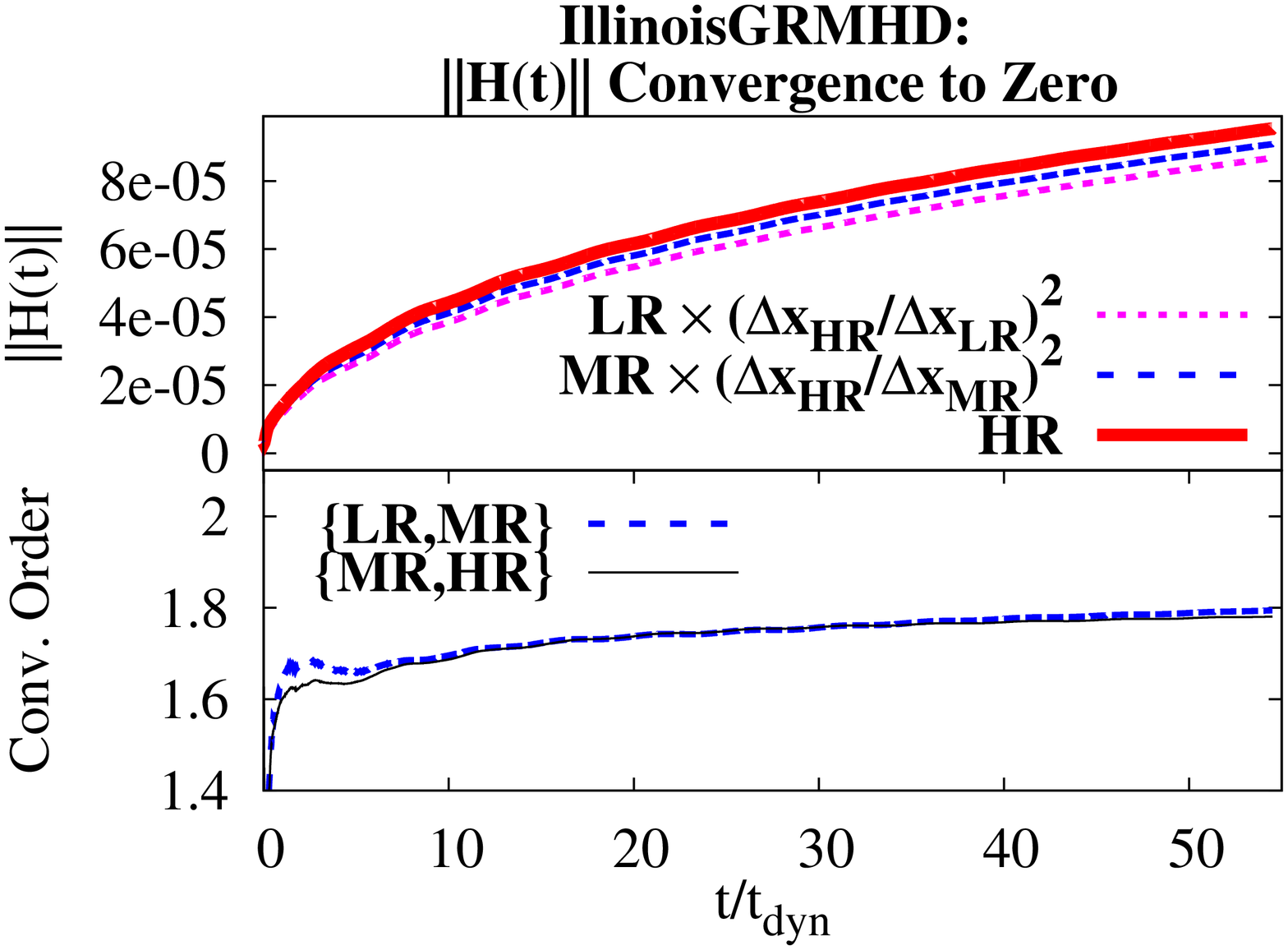}
\caption{IllinoisGRMHD and GRHydro convergence tests, for
  dynamic-spacetime, unmagnetized equilibrium TOV star evolutions
  (with physical and numerical setup as described
  in~\ref{App:Unmag_conv_test_setup}). {\bf Upper panels:} Convergence
  to zero of TOV star central density drift $\Delta \rho_c(t)
  = \rho_c(t)/\rho_c(0) - 1$, comparing GRHydro (left plot) with
  IllinoisGRMHD (right plot). Top plots show $\Delta \rho_c(t)$ at
  three separate resolutions, with the low (dotted magenta) and medium
  (dashed blue) resolution (LR and MR, respectively) simulation
  results rescaled to high resolution (HR, solid red), assuming that
  $\Delta \rho_c(t)$ converges to zero at second order. Lower plots
  show implied convergence order to zero [see Eq. \eqref{ndef}] of
  $\Delta \rho_c(t)$ for pairs of runs, for HR and MR (thin dashed
  black), and MR and LR (thick solid red), where convergence order to
  zero is defined as in 
  Eq.~\eqref{ndef}.  {\bf Lower panels:} Convergence to zero
  Eq.~\eqref{ndef} of L2 Norm of Hamiltonian constraint violation,
  $||H(t)||$. Top plots show $||H(t)||$ at three resolutions, rescaled
  so that LR (dotted magenta) and MR (dashed blue) results should
  overlap HR (solid red) results if second-order convergence to zero
  is achieved. The bottom plots show implied observed convergence
  order to zero of pairs of runs: HR and MR (dashed black), and MR and
  LR (dashed blue).  }
\label{IllinoisGRMHD_vs_GRHydro_Convergence}
\end{figure}

\begin{figure}
\raggedleft
\includegraphics[angle=270,width=0.8\textwidth]{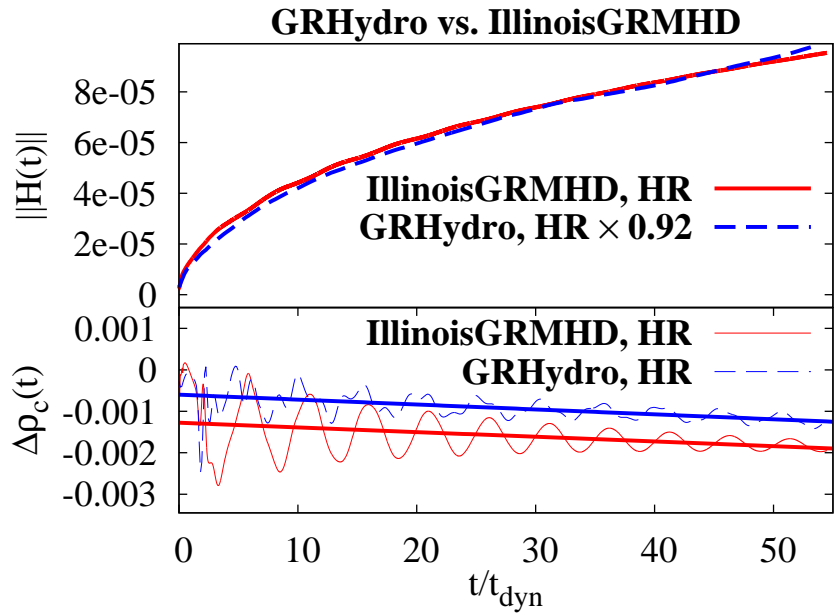}
\caption{Truncation-error analysis, comparing results from
  IllinoisGRMHD and GRHydro at high resolution (HR). The top plot
  shows L2-Norm Hamiltonian constraint violation, for IllinoisGRMHD
  (red solid) and GRHydro (blue dashed). GRHydro exhibits about 8\%
  higher constraint violation, so its data were multiplied by 0.92 to
  achieve a good overlap with IllinoisGRMHD data. The bottom plot
  shows central density drift $\Delta \rho_c(t) = \rho_c(t)/\rho_c(0)
  - 1$, at high resolution (HR) as well, for IllinoisGRMHD (thin
  dashed blue) and GRHydro (thin solid red). The thick blue and red
  lines are linear least-squares fits to IllinoisGRMHD and GRHydro
  data, from $5 t_{\rm dyn}$ to $55 t_{\rm dyn}$. The slope on the
  GRHydro line is $(-1.19 \pm 0.02) \times 10^{-5}$, and $(-1.13 \pm
  0.04) \times 10^{-5}$ for IllinoisGRMHD, where the errors given are
  standard deviations.
}
\label{IllinoisGRMHD_vs_GRHydro_HR}
\end{figure}

Figure~\ref{IllinoisGRMHD_vs_GRHydro_Convergence} demonstrates that
for this stable, equilibrium TOV star, truncation errors lead to
nonzero drifts in star's central density and the L2-Norm of the
Hamiltonian constraint, which each converge to zero at roughly
second-order [$n(t)\approx 2$, where $n$ is as defined in Eq.~\eqref{ndef}].  Notice that L2-Norm Hamiltonian
constraint convergence order fluctuates significantly in GRHydro
evolutions, as compared to IllinoisGRMHD. Additionally, at the highest
resolution chosen (resolving the NS diameter to approximately 80 gridpoints), GRHydro
possesses roughly $8\%$ higher Hamiltonian constraint violation than
IllinoisGRMHD, as shown in Fig.~\ref{IllinoisGRMHD_vs_GRHydro_HR}. We
conclude that IllinoisGRMHD appears to suffer from less Hamiltonian
constraint violation than GRHydro at a given resolution, and exhibits
more consistent L2-Norm constraint violation convergence to zero as
resolution is increased.

However, at all resolutions, the absolute value of central density
drift through 55 dynamical timescales is far higher with IllinoisGRMHD
than GRHydro. We analyze the drift at high resolution in the lower panel of
Fig.~\ref{IllinoisGRMHD_vs_GRHydro_HR}, finding that the differences
appear directly after initial settling of the TOV star. These
differences are not surprising, given the unique algorithmic choices
in each code (e.g., GRHydro adopts the internal energy $\epsilon$ as a
primitive variable, where IllinoisGRMHD adopts pressure $P$ instead,
just to name one). In this plot, we fit
data in the range $5\leq t/t_{\rm dyn} \leq 55$ to a least-squares
linear trendline, finding that the rate of central density drift
(i.e., the slope of the linear trendline) after 5 dynamical timescales
to be within about one standard deviation for the two codes. In
addition, we verified that although the simulation is run for about
2.3 light-crossing times, doubling the outer boundary has no
qualitative effect on the results.

In a forthcoming paper, we will demonstrate that differences between
these two open-source GRMHD codes spawn from how the GRMHD
evolution algorithms are {\it implemented}, independent of the chosen
reconstruction scheme.

\section{Performance Benchmarks}
\label{Bench}

We have demonstrated that although IllinoisGRMHD represents a complete
rewrite of OrigGRMHD,
the two codes agree to roundoff precision. IllinoisGRMHD is also
designed to be more user-friendly, more extensible, and better
documented than OrigGRMHD as well. Furthermore, as we demonstrate
in this section, IllinoisGRMHD performs and scales better than
OrigGRMHD. This stems from the fact that coding decisions
within IllinoisGRMHD were made specifically from the outset to optimize
not only user-friendliness and code readability, but also
performance.

Making IllinoisGRMHD perform as well as GRHydro, on the other hand,
appears to be an unlikely goal, as the AMR-capable GRMHD algorithm
adopted by IllinoisGRMHD/OrigGRMHD is far more computationally
intensive. All variables in GRHydro's GRMHD scheme for AMR grids
(hyperbolic divergence cleaning~\cite{HDC2002,GRHYDRO}) are
unstaggered, overlapping gridpoints. Meanwhile,
IllinoisGRMHD/OrigGRMHD implement a staggered vector-potential
formulation, requiring, e.g., about 60\% more 
expensive reconstructions to compute GRMHD fluxes, as they must be
computed on {\it staggered} gridpoints. In addition, the evolution of
the staggered EM vector potential gauge quantity $[\sgam \Phi]$ is
quite expensive, as it requires a large number of interpolations. Of
course, in exchange for this more expensive algorithm is the guarantee
that monopoles (i.e., violations of $\nabla \cdot \ve{B}=0$) cannot be
generated on grid refinement boundaries when magnetized fluid flows
cross them. GRHydro cannot guarantee this, but IllinoisGRMHD/OrigGRMHD
does.

\begin{figure}
\raggedleft
\includegraphics[angle=270,width=0.8\textwidth]{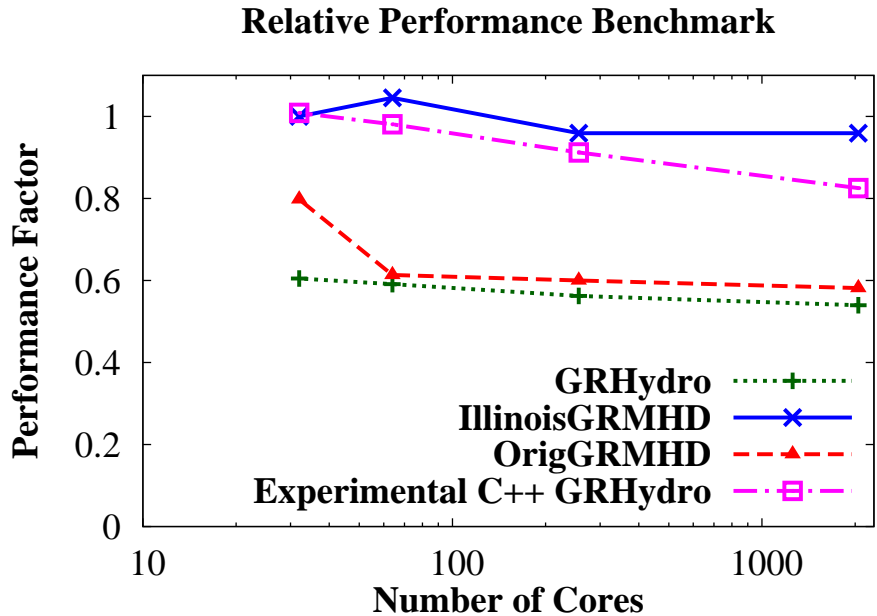}
\caption{Relative performance of IllinoisGRMHD (blue solid), OrigGRMHD
  (red dashed), standard GRHydro (green dotted), and
  the experimental C++ version of GRHydro (magenta dot-dashed) at multiple
  problem scales on the Stampede supercluster, performing an
  unmagnetized neutron star simulation, but with magnetic field
  evolution enabled. Performance data are normalized to IllinoisGRMHD
  two-node (32-core) performance (as measured by the number of
  gridzones computed per second per processor core). As the number of
  cores was increased, the number of gridpoints per core was kept
  fixed at approximately $72^3$ for all four refined levels and $68^3$
  for the lowest-resolution level on these AMR grids, effectively
  making this a weak-scaling test.}
\label{Stampede_weak_scaling}
\end{figure}

Thus {\it a priori}, we would expect IllinoisGRMHD and OrigGRMHD to
significantly under-perform GRHydro. Remarkably,
Fig.~\ref{Stampede_weak_scaling} demonstrates that for a physical
system and AMR grid hierarchy typically used in 
production runs, IllinoisGRMHD actually {\it outperforms} both GRHydro
{\it and} OrigGRMHD by a comfortable margin. There also exists a new,
experimental C++ version of GRHydro (henceforth,
GRHydro-experimental), which was written in part to improve performance. Indeed,
performance is greatly enhanced by GRHydro-experimental, but it at
best matches IllinoisGRMHD performance at small core counts, and
scales worse than IllinoisGRMHD with increasing core count. The physical
system and basic AMR grid hierarchy is as described in
\ref{App:Unmag_conv_test_setup}; i.e., it consists of an unmagnetized,
equilibrium TOV star for which the magnetic-field is also evolved but
initialized to zero.

As measured by the number of gridpoints computed per second per core
on the TACC Stampede supercomputer, at all problem scales typically
used for parallel AMR runs (ranging from 32 to 2,048 cores),
Fig.~\ref{Stampede_weak_scaling} shows that IllinoisGRMHD consistently
outperforms the standard GRHydro by a factor of between
1.7--1.8. However, IllinoisGRMHD matches GRHydro-experimental's
performance, to within measurement error at 32 cores, but manages to
outperform GRHydro-experimental by about 16\% at 2,048 cores. Again,
it is remarkable that IllinoisGRMHD can produce performance numbers in
the same ballpark as GRHydro, as IllinoisGRMHD implements a much more
expensive GRMHD algorithm. 

The performance improvement over OrigGRMHD is also significant, with
IllinoisGRMHD outperforming OrigGRMHD by a factor of 1.3 at 32 cores,
increasing to 1.6 at 256 and 2,048 cores. In independent testing, we
find that about $10-20\%$ of the performance difference between
IllinoisGRMHD and OrigGRMHD is due to the fact that OrigGRMHD is based
on an old, unmaintained version of the Cactus/Carpet infrastructure
(ca. October, 2010). For more details on the physical system and basic
grid structure of this benchmark, see~\ref{App:Unmag_conv_test_setup}.

Benchmarks presented here measure {\it total} simulation performance,
and since these are dynamical spacetime simulations, the performance
gap between IllinoisGRMHD and the other codes will certainly increase
for fixed-background-spacetime simulations. Thus by adopting
IllinoisGRMHD, research groups currently using the standard version of
GRHydro or OrigGRMHD stand to boost their computational resources by a
factor of between 1.6--1.8. Independent tests indicate that the
performance gap increases to a factor of $\approx 2$ in {\it
  fixed}-spacetime-background GRMHD simulations.

Making AMR-based codes like IllinoisGRMHD, OrigGRMHD, and GRHydro
scale well is an intrinsically difficult task. AMR greatly reduces
the memory and processor overhead in our simulations, focusing
resolution only where it is needed, and generating many small, refined
numerical grids in the process. When these small refined grids are
parallelized, however, they are generally split into even smaller
grids, resulting in a large grid surface area to volume ratio. As the
information on the surfaces must be communicated across nodes, this
makes the performance of AMR-based codes strongly network-limited.

OpenMP~\cite{OpenMP} can be used to combat this by splitting
computational loops over multiple processor cores, enabling us to use
fewer parallel (MPI) processes per CPU and thus larger grids on a
given (MPI) process. This reduces the network load significantly and
thus increases overall performance. 

\begin{figure}
\raggedleft
\includegraphics[angle=270,width=0.8\textwidth]{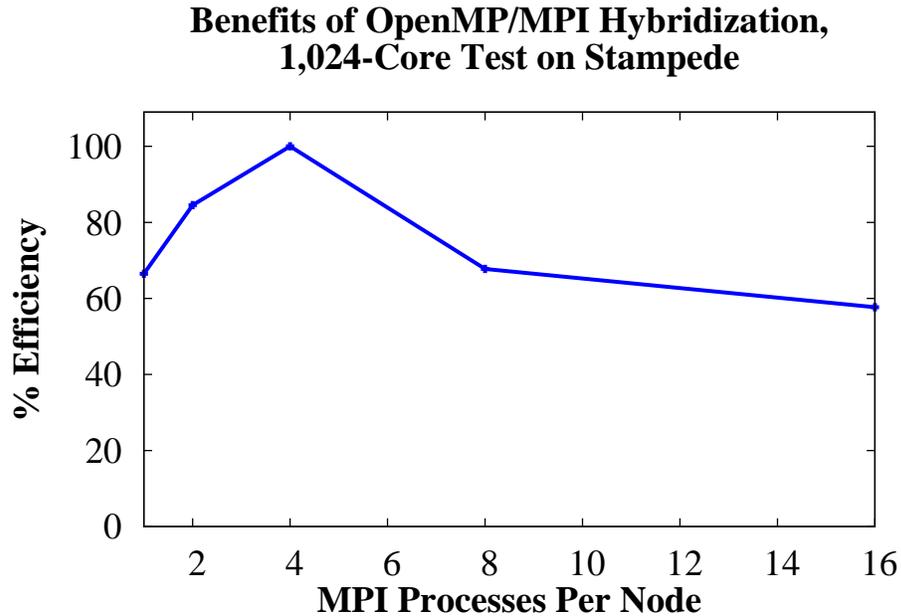}
\caption{IllinoisGRMHD code performance as the number of MPI processes
per node on Stampede is varied, with the total core-count fixed at
1,024 (i.e., 64 Stampede nodes). When running with 1,2,4,8, and 16
MPI processes per node, the number of threads per MPI process
({\tt OMP\_NUM\_THREADS}) was set to 16,8,4,2, and 1,
respectively. Efficiency is normalized to the 4 MPI processes per node
case. Simulation is of a neutron star with full GRMHD and spacetime
evolution enabled.}
\label{OpenMP_MPI_Hybrid}
\end{figure}

As shown in Fig.~\ref{OpenMP_MPI_Hybrid}, in production-scale
benchmarks, we find that IllinoisGRMHD performs
slightly more than 40\% faster as an OpenMP/MPI hybrid code than as a
pure MPI code (i.e., when running 16 MPI processes per node, OpenMP
disabled), for a typical GRMHD production run on the Stampede
supercluster. Although all three codes possess some degree of OpenMP
support, all of IllinoisGRMHD's loops have has been written with full
OpenMP~\cite{OpenMP} support, making IllinoisGRMHD a pure OpenMP/MPI
hybrid code, just like OrigGRMHD. We finish this section by noting
that all benchmark results in Fig.~\ref{Stampede_weak_scaling} were
performed using 4 MPI processes per node and 4 OpenMP threads per MPI
process, which we found maximized performance in all codes.

\section{Conclusions and Future Work}
\label{Conclusions}

The field of numerical relativity has matured considerably in the
years since the first dynamical spacetime GRMHD codes were developed,
and multiple groups now possess such codes. Given that the future of
our field depends on the ability to advance
and extend these codes to model new physics, while still maintaining
and improving the GRMHD modules, it stands to reason that the
community could benefit if we consolidated our efforts and
adopted the same dynamic-spacetime GRMHD code.

With its proven robustness and reliability in modeling some of the
most extreme phenomena in the Universe, it seems the
OrigGRMHD code could be a good candidate for such community adoption
if it were open-sourced. But despite its strong scientific track
record, OrigGRMHD was not written with community adoption in mind,
instead being a code written ``by experts and for experts'' of
OrigGRMHD, with a premium put on immediate applications. As such,
the code lacked a number of features common to large, open-source,
community-based codes in computational astrophysics, including sufficient
documentation and code comments, fine-grained modularity, a consistent
coding style, and regular, enforced code maintenance (e.g., removal of
unused and unmaintained features).

As an open-source, from-the-ground-up rewrite of OrigGRMHD,
IllinoisGRMHD aims to fix all the former code's
idiosyncracies, thus facilitating widespread community
adoption. With such adoption in mind, IllinoisGRMHD's development has
been guided by the four core design principles of
user-friendliness, modularity/extensibility, robustness, and 
performance/scalability. Regarding user-friendliness, the code is
well-documented, properly commented, and requires only basic
programming skills to understand and run. IllinoisGRMHD is also far more
modular and extensible than OrigGRMHD, with low-level CFD routines
split off from the main code into a library of extensible functions.

As for robustness, IllinoisGRMHD was designed to act as a drop-in
replacement for OrigGRMHD, and we have demonstrated that IllinoisGRMHD
indeed reproduces results from the original code to roundoff-level
precision, not only when evolving magnetized neutron stars, but also
discontinuous, random initial data.

In addition, IllinoisGRMHD largely produces consistent results with the
only other open-source, dynamical spacetime GRMHD code, GRHydro, in
that both codes exhibit approximate second-order convergence. Although
both codes were run with the same basic evolution algorithms, results
differ due to the specific details of how these algorithms were
implemented. We will explore this further in a forthcoming work, but
just to name a couple of differences, GRHydro reconstructs the
specific internal energy $\epsilon$ and the Valencia-formulation
3-velocity, while IllinoisGRMHD reconstructs pressure and 3-velocity
defined as $v^i = u^i/u^0$. When evolving equilibrium unmagnetized 
TOV stars, GRHydro produces about 8\% higher Hamiltonian constraint
violations (as measured by the L2 Norm over the entire grid), but
significantly less absolute central density drift than
IllinoisGRMHD. We find that the rate of central density drift is
identical between the two codes after the star undergoes an initial
settling over a few dynamical timescales. We conclude it is this
initial settling that causes the large discrepancy in absolute central
density drift.

Though user-friendliness, modularity/extensibility, and robustness
were the primary considerations in IllinoisGRMHD's development, it
would be hard to convince key developers of other codes to adopt
IllinoisGRMHD unless we could demonstrate at least comparable
performance and scalability to alternative dynamical spacetime GRMHD
codes. To this end, we have shown that IllinoisGRMHD is in fact about
1.7--1.8 times faster than the standard version of GRHydro for
production-size, AMR-enabled, GRMHD runs on the Stampede supercluster,
scaling to typical high-resolution core-counts at better than 95\%
efficiency. 

This is a rather remarkable result, as IllinoisGRMHD implements a far
more computationally expensive GRMHD algorithm than GRHydro to ensure
the no-monopole constraint $\nabla \cdot \ve{B}=0$ is satisfied. This
added expense forbids the generation of monopoles at
AMR grid boundaries in the case of multi-scale GRMHD flows, which
GRHydro, and even its new, experimental C++ version, which we refer to
as GRHydro-experimental, cannot guarantee.  GRHydro-experimental,
which was actively being written during the preparation of this paper,
is a complete rewrite of the standard, FORTRAN-based GRHydro, with a
goal in part of improving performance. Indeed, GRHydro-experimental does improve performance,
but only at best matches IllinoisGRMHD's performance at small core counts, with IllinoisGRMHD
performing about 16\% better at 1,024 to 2,048 cores. Though these
results may change when simulating other systems of interest, we
consider them representative.

IllinoisGRMHD also outperforms OrigGRMHD by a factor of 1.3--1.6. Thus
by adopting IllinoisGRMHD, research groups using GRHydro or OrigGRMHD
stand to increase their computational resources available for
dynamical spacetime, GRMHD runs. While in terms of performance,
IllinoisGRMHD seems to be only slightly better than the experimental,
C++ version of GRHydro, perhaps the two greatest advantages of
IllinoisGRMHD over GRHydro is that (1) IllinoisGRMHD does not allow the
generation of magnetic monopoles when modeling multi-scale GRMHD flows
on AMR grids and (2) IllinoisGRMHD is capable of stably modeling GRMHD
flows into BH horizons over very long timescales, without the need for
special algorithms that excise GRMHD data with the BH. GRHydro
requires excision to model such flows and its GRMHD features have been
mostly used for core collapse (to a neutron star) simulations, in
which no BH is present.
We conclude that making GRHydro's
GRMHD schemes as robust may require careful specification of boundary
conditions on the excision surface coupled to an interpolation scheme
across AMR level boundaries that respects the no-monopoles constraint.

As mentioned previously, a forthcoming paper will analyze how
differences in algorithmic implementations between GRHydro and
IllinoisGRMHD can lead to significantly different results when
evolving neutron stars. One possibility is that we may find an
implementation that results in a superior code to either original
code. If such a code is found, it may prove quite useful to reliably
evolving binary neutron stars and black hole--neutron stars over
many orbits with a minimum of computational expense.

Although IllinoisGRMHD is ready for production runs now, we encourage
other developers to join our effort in improving IllinoisGRMHD beyond
its current state, as a great deal of important work remains to be
done. We would like to port features from GRHydro into IllinoisGRMHD's
library of functions, using IllinoisGRMHD's coding style, including
reconstruction schemes, conservative-to-primitives solvers, and more
advanced approximate Riemann solvers, just to name a
few. IllinoisGRMHD was originally written in a standalone sandbox for
maximum portability, and was only recently ported into the Einstein
Toolkit. It therefore makes minimal use of certain aspects of the ET
infrastructure that could greatly extend its usefulness. For example,
the current version supports only single gamma-law EOSs, and full 3D
simulations with either no symmetries enabled, or simply bitant
symmetry across the $xy$-plane. The ET infrastructure provides support
for arbitrary EOSs, symmetry conditions, etc., and we intend to work
with the large ET community toward making this code the standard
choice in the ET, and one the community can be proud of. 

\ack

We gratefully acknowledge the ET community for hosting the
IllinoisGRMHD software, and Sean T.~McWilliams for useful discussions
and for generously lending time on the Spruce Knob HPC resource at
WVU. This paper was supported in part by NSF Grant PHY-1300903 and
NASA Grants NNX13AH44G and 13-ATP13-0077. VP
is supported in part by the Simons Foundation and by NSF grant
PHY-1305682. This work used the Extreme Science and Engineering
Discovery Environment (XSEDE), which is supported by NSF grant number
OCI-1053575. 

\appendix
\section{Physical and Computational Setup for Unmagnetized TOV Star
  Convergence Tests}
\label{App:Unmag_conv_test_setup}

IllinoisGRMHD and GRHydro evolve unmagnetized, polytropic TOV star
initial data, consisting of a TOV star with central density of
0.129285309 constructed with a polytropic equation of state $P=K
\rho^\Gamma$, where $K=1$ and $\Gamma=2$. This generates a model for a
cold, degenerate neutron star (NS) with compaction $M_{NS}/R_{NS}
\approx 0.1467$. The initial data are generated using the built-in
TOVSolver thorn within the ET. The convergence tests are
dynamical spacetime tests, in which IllinoisGRMHD and GRHydro are each
coupled to the McLachlan BSSN thorn. To match IllinoisGRMHD's
evolution algorithms, the ``HLLE'' approximate Riemann solver is
chosen for GRHydro evolutions, along with PPM reconstruction. Though
the codes differ in their usage of these algorithms (e.g., GRHydro
reconstructs the internal energy $\epsilon$ instead of pressure $P$,
etc.), the parameters are chosen so that the both codes share
precisely the same algorithms. Note that the outer boundary conditions
for the spacetime variables are set to be identical between the two
codes, but the hydrodynamic boundary conditions differ. However, this
is inconsequential, as the density near the outer boundary remains
within $\approx 1\%$ of the original atmosphere value in both codes
throughout the entire evolution.

These tests are performed on cubic AMR grids, so that the coarsest grid cube
possesses a half-side-length of 40$R_{\rm NS}$, centered on the
NS. The AMR hierarchy---nested within this coarse grid and also
centered on the NS---consists of four, progressively higher-resolution
cubes with half-side-lengths of 15$R_{\rm NS}$, 7.5$R_{\rm NS}$,
3.75$R_{\rm NS}$, and 1.875$R_{\rm NS}$. At each finer level, the grid
spacing is halved, so that the cube with half-side-length 1.875$R_{\rm NS}$
has a grid spacing of $\Delta x = \{0.03906,0.03125,0.025\}R_{\rm NS}$
for low, medium, and high-resolution runs, respectively. This
corresponds to resolving the NS across its diameter to approximately
51, 64, and 80 gridpoints, for low, medium, and high-resolution runs,
respectively. In both codes, low-density atmosphere density floor is
set to correspond to $10^{-7}$ times the initial central density of
the NS.

\section{Physical and Computational Setup for Magnetized TOV Star
  Code Validation Tests}
\label{App:Mag_roundoff_test_setup}

The magnetized TOV star used in these tests is precisely the same
stable neutron star as described in~\ref{App:Unmag_conv_test_setup},
except it is seeded with a weak magnetic field at $t=0$.
This initial magnetic field is purely poloidal, with vector potential
components defined as
\beqn
A_x &=& -y A_b \max(P-P_{\rm cut},0), \\
A_y &=&  x A_b \max(P-P_{\rm cut},0), \\
A_z, [\sgam \Phi] &=& 0,
\eeqn
where $P_{\rm cut}$ is set to 4\% of the maximum pressure and $A_b$ is
set so that the maximum initial magnetic-to-gas pressure ratio
$\beta^{-1}$ is $0.83 \times 10^{-3}$. Like the convergence tests,
these tests are dynamical spacetime tests, but unlike the convergence
tests, IllinoisGRMHD and OrigGRMHD codes are each coupled to the
(closed-source) BSSN dynamical spacetime module of the Illinois NR
group. Further, the initial TOV star density, pressure, and spacetime
metric profiles are provided by the code of \cite{Cook1994}, which
generates data for this nonrotating star that agrees to machine
precision with the TOVSolver code used in the unmagnetized tests.

These tests are also performed on AMR grids, with the coarsest grid
cube extended so its half-side-length is 10$R_{\rm NS}$, centered on
the NS. Two finer AMR grid levels are nested within this coarse grid:
the next finer grid having half-side-length 5$R_{\rm NS}$ and the
finest having 2.5 $R_{\rm NS}$. Again, at each finer level, the grid
spacing is halved, so that the cube with half-side-length 2.5$R_{\rm NS}$
has a grid spacing of $\Delta x \approx \{0.078\}R_{\rm NS}$. This
corresponds to resolving the NS to 26 gridpoints across its diameter.
The resolution is intentionally set to be very low, to guarantee that
truncation-error differences between the codes will be more strongly
magnified than at (higher) resolutions typically chosen for evolving
NSs. In addition, a close outer boundary is chosen to maximize its
influence on the evolution, to check for errors in coding the
outflow outer boundary conditions. Finally, we choose our
low-density atmosphere density floor to correspond to $10^{-7}$ times
the initial central density of the NS.

\section{Random Initial Data}
\label{App:random_id}

The algorithm used within our random initial data module is as follows. At a
given gridpoint, the random number generator is first seeded with a
unique integer based on the coordinate of the gridpoint. The seed is
then used to generate one double-precision random number $\xi \in
(-0.1,0.1)$. Because this random number is based entirely on the
coordinate of a gridpoint, the initial data will be consistent when
evolving on identical grids, regardless of how the global grid is
split among processors when generating and evolving initial data in a
parallel run.

Using this double-precision random number $\xi \in (-0.1,0.1)$, we
define 
\beq
\mathcal{S} = \xi + \frac{1}{7} \left[ 4 + \sin(10 x/L) + \sin(10 y/L) + \cos(10 z/L) \right],
\eeq
where $L$ denotes the coarsest AMR grid cube half-side-length.
Thus $\mathcal{S}\in(\frac{3}{70},1.1)$ consists of a strong
stochastic perturbation atop a smoothly oscillating function. Given
the quantities $\mathcal{S}$ and $\xi$, all basic spacetime metric and
GRMHD quantities may then be set following the prescription detailed
in Table~\ref{tab:random_id_parameters}. 

\begin{table}
\begin{center}
\caption{Prescription for setting GRMHD and spacetime metric
  quantities in random initial data module. Note that the initial data
  employ the polytropic EOS $P=\rho_0^2$, and we evolve with the
  gamma-law EOS $P=(\Gamma-1)\rho_0 \epsilon$ with $\Gamma=2$.}
\begin{tabular}{cc}
  \hline
  Variable & Value set in random initial data module \\
  \hline
  $\rho_0$ & $0.01\mathcal{S}$ \\
  $P$ & $\rho_0^2$ (Gamma-law EOS) \\
  $\phi$ & $0.1\mathcal{S}$ \\
  $\alpha$ & $1 - 0.1 \mathcal{S}$ \\
  $\{v^x,v^y,v^z\}$ & $\{1,1.1,0.9\} \times 10^{-1} \xi$ \\
  $\{[\sgam \Phi],A_x,A_y,A_z\}$ & $\{0.6,1,1.1,0.9\} \times 10^{-1} \xi$ \\
  $\{\beta^x,\beta^y,\beta^z\}$ & $\{0.9,1.1,1\} \times 10^{-1} \xi$ \\
  $\{\tilde{\gamma}_{xx},\tilde{\gamma}_{yy},\tilde{\gamma}_{zz}\}$ & $1+\{10,1,10\} \times 10^{-3} \xi$ \\
  $\{\tilde{\gamma}_{xy},\tilde{\gamma}_{xz},\tilde{\gamma}_{yz}\}$ & $\{1,10,1\} \times 10^{-4} \xi$ \\
  $\{K_{xx},K_{xy},K_{xz},K_{yy},K_{yz},K_{zz}\}$ & $\{10,2,3,40,5,60\} \times 10^{-3} \xi$ \\
  \hline
\end{tabular}
\label{tab:random_id_parameters}
\end{center}
\end{table}

Although the resulting data are not designed to satisfy Einstein's
equations, be conformally flat, or be consistent with the BSSN
formalism, we make sure the spacetime metric is Lorentzian and the
three-metric positive definite. We also enforce the BSSN constraint
$\tilde{\gamma}=1$ as follows. Immediately after the quantities in
Table~\ref{tab:random_id_parameters} have been set at a given
gridpoint, we compute the resulting non-unit determinant
$\tilde{\gamma}$ and then multiply the components of
$\tilde{\gamma}_{ij}$ by $\tilde{\gamma}^{-1/3}$. In this way, the
unit determinant of the three-metric is enforced. After applying this
constraint, all GRMHD and spacetime metric quantities not specified in
Table~\ref{tab:random_id_parameters} (e.g., GRMHD conservative
variables, $B^i$, $\tilde{\gamma}^{ij}$, etc.) are directly computed
from quantities in that table.

In tests adopting this random initial data module, we choose an AMR
grid with one refinement level centered at the origin. The coarser
grid, with grid spacing 1.0, has cube half-side-length of 10, and the
finer grid, with grid spacing 0.5, has grid half-side-length of 2. Spacetime
evolution modules, as well as prolongation and restriction operations
on spacetime variables, are disabled. 

The initial maximum rest-mass density is chosen to be $\sim 10\%$
the rest-mass density of the TOV star used in other tests. The
atmospheric density floor is set to $10^{-6}$, which
is about $10^{-4}$ times the maximum possible initial density,
yielding stochastic fluctuations in density spanning about four orders
of magnitude.  Additionally, the magnetic-to-gas-pressure ratio at
$t=0$ ranges from $\sim 10^{-3}$ (gas-pressure dominated) to about
$10$ (magnetic-pressure dominated) initially. Given that initial data
at a given gridpoint are independently and randomly specified,
we conclude that physical quantities at neighboring gridpoints can
differ by several orders of magnitude, making this a very harsh
test. Despite the lack of coherent GRMHD flows, throughout the
evolution of these random initial data, the Lorentz factor limit of
$W=10$ is exceeded and subsequently enforced about 400 times. 

\section*{References}

\bibliographystyle{plain}
\bibliography{paper}

\end{document}